\documentclass[11pt,onecolumn,draft]{IEEEtran}

\usepackage{epsf,psfrag,amssymb,amsfonts,cite,amsmath,eufrak,yfonts,cases,subeqnarray}
\newtheorem{theorem}{Theorem}
\newtheorem{example}{Example}
\newtheorem{remark}{Remark}

\newtheorem{lemma}{Lemma}
\newtheorem{definition}{Definition}

\newtheorem{proposition}{Proposition}

\def\psfancypar#1#2{\begingroup\def\par{\endgraf\endgroup\lineskiplimit=0pt}
               \setbox2=\hbox{\large\sc #2}
               \newdimen\tmpht \tmpht \ht2 \advance\tmpht by \baselineskip
               \font\hhuge=Times-Bold at \tmpht
               \setbox1=\hbox{{\hhuge #1}}
               \count7=\tmpht \count8=\ht1
               \divide\count8 by 1000 \divide\count7 by \count8
               \tmpht=.001\tmpht\multiply\tmpht by \count7
               \font\hhuge=Times-Bold at \tmpht
               \setbox1=\hbox{{\hhuge #1}}
               \noindent
                \hangindent1.05\wd1
               \hangafter=-2 {\hskip-\hangindent
               \lower1\ht1\hbox{\raise1.0\ht2\copy1}%
                \kern-0\wd1}\copy2\lineskiplimit=-1000pt}

\newenvironment{ap}{
\renewcommand{\thesection}{\mbox{Appendix }\Alph{section}}
\renewcommand{\thesubsection}{\Alph{section}.\arabic{subsection
}}
\renewcommand{\theequation}{\mbox{\Alph{section}.}\arabic{equation}}
}{
\renewcommand{\thesection}{\arabic{section}}
\renewcommand{\thesubsection}{\thesection.\arabic{subsection}}
\renewcommand{\theequation}{\arabic{equation}}
}

\newcommand{\bsq}{\begin{subequations}\begin{eqnarray}}
\newcommand{\esq}{\end{eqnarray}\end{subequations}}
\newcommand{\beq}{\begin{equation}}
\newcommand{\eeq}{\end{equation}}
\newcommand{\bqa}{\begin{eqnarray}}
\newcommand{\eqa}{\end{eqnarray}}
\newcommand{\bqn}{\begin{eqnarray*}}
\newcommand{\eqn}{\end{eqnarray*}}
\newcommand{\nn}{\nonumber}

\newcommand{\be}{\begin{enumerate}}
\newcommand{\ee}{\end{enumerate}}
\newcommand{\bi}{\begin{itemize}}
\newcommand{\ei}{\end{itemize}}
\newcommand{\bd}{\begin{description}}
\newcommand{\ed}{\end{description}}
\newcommand{\ba}{\begin{array}}
\newcommand{\ea}{\end{array}}
\newcommand{\bde}{\begin{definition}}
\newcommand{\ede}{\end{definition}}
\newcommand{\bex}{\begin{example}}
\newcommand{\eex}{\end{example}}

\newcommand{\Phibf}{\mbox{${\bf \Phi}$}}

\newcommand{\abf}{\mbox{${\bf a}$}}

\def\sign{\textrm{sign}}
\def\boxit#1{\vbox{\hrule\hbox{\vrule\kern3pt
        \vbox{\kern3pt#1\kern3pt}\kern3pt\vrule}\hrule}}

\def\reals{ { {\rm  I \kern-0.15em R }  } }
\def\complex{ {\,{{\rm C} \kern-0.50em \raise0.20ex {  |}}\, }}

\def\rhobf{\hbox{\boldmath$\rho$\unboldmath}}

\def\Sigmabf{\hbox{$\bf \Sigma$}}

\def\Sigmabf{\mbox{$ \bf \Sigma $}}

\def\0bf{{\bf 0}}
\def\1bf{{\bf 1}}
\def\2bf{{\bf 2}}
\def\3bf{{\bf 3}}
\def\4bf{{\bf 4}}
\def\5bf{{\bf 5}}
\def\6bf{{\bf 6}}
\def\7bf{{\bf 7}}
\def\8bf{{\bf 8}}
\def\9bf{{\bf 9}}
\def\abf{{\bf a}}

\def\Abf{{\bf A}}
\def\Bbf{{\bf B}}
\def\Cbf{{\bf C}}
\def\Dbf{{\bf D}}
\def\Ebf{{\bf E}}
\def\Fbf{{\bf F}}

\def\Hbf{{\bf H}}
\def\Ibf{{\bf I}}

\def\Kbf{{\bf K}}

\def\Obf{{\bf O}}

\def\Rbf{{\bf R}}
\def\Sbf{{\bf S}}

\def\Ubf{{\bf U}}
\def\Vbf{{\bf V}}
\def\Wbf{{\bf W}}
\def\Xbf{{\bf X}}

\def\fp{{\pmb f}}

\def\hp{{\pmb h}}

\def\np{{\pmb n}}

\def\xp{{\pmb x}}
\def\yp{{\pmb y}}
\def\zp{{\pmb z}}

\def\Nmat{\mathcal{N}}

\def\Rmat{\mathcal{R}}

\def\Xmat{\mathcal{X}}
\def\Ymat{\mathcal{Y}}

\def\QED{\mbox{\rule[0pt]{1.3ex}{1.3ex}}}
\def\bpf{{\em Proof: }}
\def\epf{\hspace*{\fill}~\QED\par\endtrivlist\unskip}

\def\Rxx{\Rbf_{\ssstyle X\kern-.1em X}}

\let\ssstyle=\scriptscriptstyle

\def\Cov{{\textrm{Cov}}}

\def\tr{{\textrm{tr}}}
\def\Vec{{\textrm{Vec}}}
\def\rank{{\textrm{rank}}}
\def\diag{{\textrm{diag}}}

\def\arg{\textrm{arg}}

\def\Kout{\setbox1=\hbox{\Huge\bf K}\hbox to
1.05\wd1{\hspace{.05\wd1}
}}
\def\Sout{\setbox1=\hbox{\Huge\bf S}\hbox to 1.05\wd1{\hspace{.05\wd1}
}}

\def\scalefig#1{\epsfxsize #1\textwidth}
 \allowdisplaybreaks[4]

\begin{document}
\title{Capacity Region of Vector Gaussian Interference Channels with Generally Strong Interference}
\author{Xiaohu Shang, and H. Vincent Poor\thanks{X.
Shang is with Bell-Labs, Alcatel-Lucent, 791 Holmdel Rd., R-127, Holmdel, NJ, 07733. Email:xiaohu.shang@alcatel-lucent.com.
H. V. Poor is with Princeton University, Department of Electrical Engineering, Princeton, NJ, 08544. Email: poor@princeton.edu.
H. V. Poor was supported in part by the National
Science Foundation under Grant CNS-09-05398.}} \maketitle

\begin{abstract}
An interference channel is said to have strong interference if for all input distributions, the receivers can fully decode the interference. This definition of strong interference applies to discrete memoryless, scalar and vector Gaussian interference channels. However, there exist vector Gaussian interference channels that may not satisfy the strong interference condition but for which the capacity can still be achieved by jointly decoding the signal and the interference. This kind of interference is called generally strong interference. Sufficient conditions for a vector Gaussian interference channel to have generally strong interference are derived. The sum-rate capacity and the boundary points of the capacity region are also determined.
\end{abstract}

\section{Introduction}
A discrete memoryless interference channel (IC) is a quintuplet $\left(\Xmat_1,\Xmat_2,p,\Ymat_1,\Ymat_2\right)$ where $\Xmat_1$ and $\Xmat_2$ are the input alphabet sets; $\Ymat_1$, and $\Ymat_2$ are the output alphabet sets; and $p$ is a collection of conditional channel probabilities $p\left(y_1y_2\left|x_1x_2\right.\right)$ of $(y_1,y_2)\in\Ymat_1\times\Ymat_2$ given $(x_1,x_2)\in\Xmat_1\times\Xmat_2$. The receiver $i$, $i=1,2$, is required to decode $X_i$ from the received signal $Y_i$. The capacity region of this channel is known for the strong interference case \cite{Costa&ElGamal:87IT}:
\bsq
0\leq R_1&{}\leq{}& I\left(X_1;Y_1|X_2Q\right)
\label{eq:C1}\\
0\leq R_2&{}\leq{}& I\left(X_2;Y_2|X_2Q\right)
\label{eq:C2}\\
R_1+R_2&{}\leq{}& \min\left\{I\left(X_1X_2;Y_1|Q\right),I\left(X_1X_2;Y_2|Q\right)\right\}
\label{eq:Cs}
\esq
where $Q$ is a time sharing random variable. The strong interference conditions are that
\bqa
I\left(X_1;Y_1|X_2\right)\leq I\left(X_1;Y_2|X_2\right)
\label{eq:strong1}\\
I\left(X_2;Y_2|X_1\right)\leq I\left(X_2;Y_1|X_1\right)
\label{eq:strong2}
\eqa
are satisfied for all product distributions on $\Xmat_1\times\Xmat_2$.

This definition of strong interference also applies to the scalar Gaussian ICs defined in the standard form as
\bqn
Y_1&{}={}&X_1+\sqrt{a_2}X_2+Z_1\\
Y_2&{}={}&X_2+\sqrt{a_1}X_1+Z_2
\eqn
where $X_i$ and $Y_i$ $i=1,2$, are respectively the transmitted and received signals for user $i$, $Z_i$ is unit variance Gaussian noise, and $a_i$ is the cross channel gain known at both transmitters and receivers. In addition, $X_i$ has a power constraint $P_i$. The capacity region of this channel with strong interference is given in \cite{Han&Kobayashi:81IT} and \cite{Sato:81IT}:
\bqn
0\leq R_1&{}\leq{}&\frac{1}{2}\log(1+P_1)\nn\\
0\leq R_2&{}\leq{}&\frac{1}{2}\log(1+P_2)\nn\\
R_1+R_2&{}\leq{}&\min\left\{\frac{1}{2}\log(1+P_1+a_2P_2),\frac{1}{2}\log(1+P_2+a_1P_1)\right\}.
\eqn
The strong interference conditions here are
\bqa
a_1\geq 1 \quad\textrm{and}\quad a_2\geq 1.
\label{eq:scalstrong}
\eqa
It is easy to show that under the above conditions, both (\ref{eq:strong1}) and (\ref{eq:strong2}) hold for all distributions of $X_1$ and $X_2$. Therefore, the strong interference conditions for the scalar Gaussian IC coincide with those for  the discrete memoryless IC.

\begin{figure}[h] \centerline{
\begin{psfrags}
\psfrag{x1}[c]{$\xp_1$} \psfrag{x2}[c]{$\xp_2$}
\psfrag{y1}[c]{$\yp_1$} \psfrag{y2}[c]{$\yp_2$}
\psfrag{n1}[c]{$\zp_1$} \psfrag{n2}[c]{$\zp_2$} \psfrag{+}[c]{$+$}
\psfrag{g11}[c]{$\Hbf_1$} \psfrag{g12}[c]{$\Fbf_1$}
\psfrag{g21}[c]{$\Fbf_2$} \psfrag{g22}[c]{$\Hbf_2$}
\scalefig{.35}\epsfbox{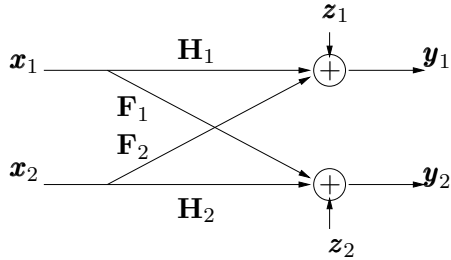}
\end{psfrags}}
\caption{\label{fig:model} The two-user MIMO IC.}
\end{figure}

Since the capacity region was determined for scalar Gaussian ICs under strong interference, substantial effort has been devoted to extending the strong interference conditions to the
multiple-input multiple-output (MIMO) IC. As shown in Fig.
\ref{fig:model}, the received signals for a MIMO IC are defined as
 \bqa
&&\hspace{-.2in}\yp_1=\Hbf_1\xp_1+\Fbf_2\xp_2+\zp_1\nn\\
&&\hspace{-.2in}\yp_2=\Hbf_2\xp_2+\Fbf_1\xp_1+\zp_2%
\label{eq:model}%
\eqa
where $\xp_i,i=1,2,$ is the transmitted (column) vector signal of user $i$ which
is subject to the average power constraint
\bqa%
\sum_{j=1}^n\tr\left(E\left[\xp_{ij}\xp_{ij}^T\right]\right)\leq nP_i
\label{eq:CovConstraint}
\eqa%
where $\xp_{i1},\xp_{i2},\ldots,\xp_{in}$, is the transmitted vector
sequence of user $i$, and $P_i$ is the power constraint.
The noise $\zp_i$ is a
Gaussian random vector with zero mean and identity covariance
matrix; and $\Hbf_{i}$ and $\Fbf_i$, $i=1,2$, are the channel
matrices known at both the transmitters and receivers. Transmitter $i$ has $t_i$ antennas and receiver $i$ has $r_i$ antennas. Without loss of generality, we assume $\Hbf_i\neq\0bf$ and $P_i>0$.

In \cite{Vishwanath&Jafar:04ITW}, the capacity region of a single-input-multiple-output (SIMO) IC with strong interference was determined. In this SIMO IC, the channel matrices are $\Hbf_i=\hp_i$ and $\Fbf_i=\fp_i$ where $\hp_i$ and $\fp_i$ are column vectors. A SIMO IC is said to have strong interference if
\bqa
0<\|\hp_i\|\leq\|\fp_i\|,\quad i=1,2\nn
\eqa
where $\|\cdot\|$ is the Euclidian vector norm.

The capacity region of a MIMO IC with strong interference was determined in \cite{Shang-etal:10IT_mimo}. A MIMO IC is said to have strong interference if there exists matrices $\Abf_i$ such that
\bqa
&&\Hbf_i=\Abf_i\Fbf_i
\label{eq:mimos1}\\
&&\Abf_i\Abf_i^T\preceq\Ibf
\label{eq:mimos2}
\eqa
for $i=1,2$, where $\Ibf$ is an identity matrix, $\Abf_i^T$ is the transpose of $\Abf_i$, and $\Abf\succeq\Bbf$ means that $\Abf$, $\Bbf$ and $\Abf-\Bbf$ are all symmetric positive semi-definite. It can be shown that if $\Hbf_i=\hp_i$ and $\Fbf_i=\fp_i$, then we can choose $\Abf_i=\hp_i\left(\fp_i^T\fp_i\right)^{-1}\fp^T_i$ and (\ref{eq:mimos2}) reduces to $\|\hp_i\|\leq\|\fp_i\|$. Therefore, the strong interference condition in \cite{Shang-etal:10IT_mimo} includes that in \cite{Vishwanath&Jafar:04ITW} as a special case. Since under condition (\ref{eq:mimos1}) and (\ref{eq:mimos2}), one can show that (\ref{eq:strong1}) and (\ref{eq:strong2}) are always satisfied, the strong interference conditions for the MIMO IC, like the scalar Gaussian IC, coincide with those for the discrete memoryless IC.

The coincidence of the strong interference conditions for discrete memoryless ICs, scalar ICs and MIMO ICs seems to have captured the essence of the IC with strong interference. All these channels have the same capacity achieving coding scheme and the same expression for the capacity region.
However, there are still observations which lead us to reconsider the strong interference condition.

To elaborate, we first introduce the concept of very strong interference \cite{Sato:78IT_strong}. A discrete memoryless IC is said to have very strong interference if
\bqa
I\left(X_1;Y_1|X_2\right)\leq I\left(X_1;Y_2\right)
\label{eq:dmcVS1}\\
I\left(X_2;Y_2|X_1\right)\leq I\left(X_2;Y_1\right)
\label{eq:dmcVS2}
\eqa
are satisfied for all product distributions on $\Xmat_1\times\Xmat_2$. Obviously, the very strong interference condition is a special case of the strong interference condition. The capacity region of such a channel is also given in (\ref{eq:C1})-(\ref{eq:Cs}) where $(\ref{eq:Cs})$ becomes inactive. However, the application of (\ref{eq:dmcVS1}) and (\ref{eq:dmcVS2}) to Gaussian ICs becomes very difficult. We use instead
\bqn
a_i\geq 1+P_i,\quad i=1,2
\eqn
as the very strong interference condition for the scalar Gaussian IC; and use
\bqa
&&\log\left|\Ibf+\Hbf_i\Sbf_i^o\Hbf_i^T\right|\leq \log\left|\Ibf+\Fbf_i\Sbf_i^o\Fbf_i^T\left(\Ibf+\Hbf_j\Sbf_j^o\Hbf_j^T\right)^{-1}\right|,\quad
i,j\in\{1,2\}, i\neq j\nn
\eqa
where
\bqa
&&\Sbf_i^o=\arg\max_{\tr(\Sbf_i)\leq P_i,\Sbf_i\succeq\0bf}\left\{\left|\Ibf+\Hbf_i\Sbf_i^o\Hbf_i^T\right|\right\}\nn
\eqa
as the very strong interference condition for the MIMO IC \cite{Shang-etal:10IT_mimo,Shang-etal:09Asilomar}. In both the scalar and MIMO ICs, the very strong interference condition can be generalized into
\bqa
I\left(\xp_i^o;\yp_i\left|\xp_j^o\right.\right)\leq I\left(\xp_i^o;\yp_j\right),\quad
i,j\in\{1,2\}, i\neq j
\label{eq:Gvstrong}
\eqa
where
\bqa
p\left(\xp_i^o\right)=\arg\max_{p\left(\xp_i\right)}I\left(\xp_i;\yp_i\left|\xp_j\right.\right).
\label{eq:gvsdist}
\eqa
Or equivalently, a Gaussian IC is said to have very strong interference if its capacity region is
\bqa
0\leq R_i\leq \max_{p\left(x_i\right)} I\left(X_i;Y_i|X_j\right),\quad i,j\in\{1,2\}, i\neq j.
\label{eq:vstrong}
\eqa
For the new very strong interference condition, the original requirement of inequalities (\ref{eq:dmcVS1}) and (\ref{eq:dmcVS2}) being satisfied for all input distributions has been relaxed to only the special input distribution (\ref{eq:gvsdist}). Clearly, the new definition includes the old one as a special case, i.e., all the ICs that satisfy (\ref{eq:dmcVS1}) and (\ref{eq:dmcVS2}) must also satisfy (\ref{eq:Gvstrong}). Although, in both cases, the capacity region is achieved by decoding the interference before the useful signal, condition (\ref{eq:gvsdist}) considers only the capacity achieving input distribution instead of all possible input distributions.

In adapting the very strong interference condition from the discrete memoryless IC to the Gaussian IC, necessary changes have been made to make it more appropriate. Comparing the very strong interference condition (\ref{eq:Gvstrong}) and the strong interference conditions (\ref{eq:strong1}) and (\ref{eq:strong2}) or (\ref{eq:mimos1}) and (\ref{eq:mimos2}), we can see some inconsistency:
\be
\item For the scalar Gaussian IC, the very strong interference condition ($a_i\geq 1+P_j$) is a special case of the strong interference condition $a_i\geq 1$. However, for the MIMO IC the very strong interference condition is generally not a special case of strong interference. As an example, we consider a MIMO IC with
    \bqn
    \Hbf_1=\Hbf_2=\left[\begin{array}{cc}
                          1 & \quad 0 \\
                          0 & \quad 1
                        \end{array}
    \right],\quad \Fbf_1=\Fbf_2=\left[\begin{array}{cc}
                          0.8 & \quad 0 \\
                          0 & \quad 2
                        \end{array}
    \right],\quad P_1=P_2=2.
    \eqn
    This MIMO IC has very strong interference (\ref{eq:Gvstrong}), and its capacity region is (\ref{eq:vstrong}). However, the strong interference conditions (\ref{eq:mimos1}) and (\ref{eq:mimos2}) are violated. Similar examples for the MIMO Z interference channel (ZIC) can be found in \cite[example 1]{Shang-etal:09Asilomar}, and examples for the MIMO IC with covariance constraints can be found in \cite[example 1]{Shang-etal:10IT_mimo}.
\item There exist many MIMO ICs for which even the matrix $\Abf_i$ in (\ref{eq:mimos1}) does not exist. For example, the multiple-input-single-output (MISO) IC: $\Hbf_i=\hp_i^T$ and $\Fbf_i=\fp_i^T$, where $\hp_i$ and $\fp_i$ are column vectors. If $\hp_i$ and $\fp_i$ are linearly independent, then the $\Abf_i$ (now a scalar) in (\ref{eq:mimos1}) does not exist. Moreover, conditions (\ref{eq:strong1}) and (\ref{eq:strong2}) are also violated if user $i$ implements zero-forcing beamforming: $I\left(X_1;Y_1|X_2\right)>0=I\left(X_1;Y_2|X_2\right)$ and $I\left(X_2;Y_2|X_1\right)>0=I\left(X_2;Y_1|X_1\right)$. However, one can still find examples for MISO ICs that have very strong interference.
\item Even for the discrete memoryless IC, there are examples which have very strong interference in the sense of (\ref{eq:Gvstrong}) instead of (\ref{eq:dmcVS1}) and (\ref{eq:dmcVS2}), and do not have strong interference \cite[section IV-B]{Xu-etal:10ITW}.
\ee

The above inconsistencies motivate us to reconsider whether there are more appropriate strong interference conditions than those in \cite{Han&Kobayashi:81IT,Sato:81IT,Costa&ElGamal:87IT} and \cite{Shang-etal:10IT_mimo} for MIMO ICs:
\be
\item The very strong interference condition requires only the capacity achieving distribution to satisfy (\ref{eq:Gvstrong}). On the contrary, the strong interference condition requires all possible input distributions to satisfy (\ref{eq:strong1}) and (\ref{eq:strong2}) or (\ref{eq:mimos1}) and (\ref{eq:mimos2}). This is generally unnecessary since we are interested in only the capacity achieving distributions. The rates achieved by other input distributions are all superseded by those achieved by the capacity achieving input distributions.
\item If  (\ref{eq:strong1}) and (\ref{eq:strong2}) hold for any input distribution, the strong interference capacity region for a discrete memoryless IC can be written as
\bqa
0\leq R_1&{}\leq{}& \min\left\{I\left(X_1;Y_1|X_2Q\right),I\left(X_1;Y_2|X_2Q\right)\right\}\nn\\
0\leq R_2&{}\leq{}& \min\left(I\left(X_2;Y_2|X_2Q\right),I\left(X_2;Y_1|X_1Q\right)\right\}\nn\\
R_1+R_2&{}\leq{}& \min\left\{I\left(X_1X_2;Y_1|Q\right),I\left(X_1X_2;Y_2|Q\right)\right\}.
\label{eq:compundMAC}
\eqa
The above region is actually the same as the capacity region of the compound multiple access channel, in which both receivers are required to {\em correctly} decode messages from both transmitters. However, for an IC any error incurred when user $i$ is trying to decode user $j$'s message, $j\neq i$, does not contribute to its overall error probability. In fact, we will show later in Lemma \ref{lemma:inner} that the rate region given in (\ref{eq:C1})-(\ref{eq:Cs}) is achieved exactly by requiring user $i$ to jointly decode $X_i$ and $X_j$.\footnote{Here `jointly decoding' means that user $i$ recovers the message from transmitter $i$ by searching the jointly typical sequence set $A_\epsilon^{(n)}\left(X_iX_jY_i\right)$. User $i$ is required to correctly decode the message from transmitter $i$. However, whether user $i$ can correctly decode the message from transmitter $j$ is not important. See the proof of Lemma \ref{lemma:inner} for further details.} Therefore, the key is whether or not joint decoding can achieve the capacity. Even though the condition that (\ref{eq:strong1}) and (\ref{eq:strong2}) hold for any input distribution is crucial in deriving the strong interference capacity region in \cite{Costa&ElGamal:87IT} and \cite{Han&Kobayashi:81IT}, these two conditions are in general not necessary conditions for joint decoding to achieve the capacity region.
\ee

Therefore, we define a new strong interference condition as follows:
\begin{definition}
An IC is said to have generally strong interference, if its capacity region is given by (\ref{eq:C1})-(\ref{eq:Cs}); or equivalently, if the capacity region is achieved by jointly decoding the signal and the interference at each receiver.
\end{definition}

In this new definition, as long as joint decoding achieves the capacity, the IC is said to have generally strong interference. Thus, we focus on only the input distribution and the coding scheme that achieve the boundary of the capacity region, instead of any possible input distributions. For the IC with generally strong interference, there may exist input distributions such that the receiver cannot correctly decode the signal and the interference.

There are cases in which only part of the boundary of the capacity region is characterized by (\ref{eq:C1})-(\ref{eq:Cs}), i.e., the IC may have generally strong interference at some rates and not at other rates (see Example \ref{example:partial} in which partially decoding the interference outperforms jointly decoding the signal and interference at some rates). Therefore, we define:
\begin{definition}
An IC is said to have generally strong interference sum-rate capacity, if its sum-rate capacity is given by the maximum sum-rate of region (\ref{eq:C1})-(\ref{eq:Cs}); or equivalently, if the sum-rate capacity is achieved by jointly decoding the signal and the interference at each receiver.
\end{definition}
\begin{definition}
An IC is said to have generally strong interference at $\{R_1,R_2\}$, if $\{R_1,R_2\}$ is on the boundary of the capacity region and satisfies (\ref{eq:C1})-(\ref{eq:Cs}) for some input distributions of $X_1$ and $X_2$; or equivalently, if $\{R_1,R_2\}$ is achieved by jointly decoding the signal and the interference at each receiver.
\end{definition}

In this paper, we study the capacity region of MIMO ICs with generally strong interference. Clearly, the generally strong interference condition includes strong interference, as well as very strong interference, as special cases. 

The rest of the paper is organized as follows: in Section II, we derive sufficient conditions for a MIMO IC to have generally strong interference by comparing an inner bound and an outer bound for the capacity region; in Sections III and IV, we apply these sufficient conditions to SIMO and MISO ICs respectively, and obtain simplified generally strong interference conditions; numerical examples are given in Section V; and we conclude in Section VI.

Before proceeding, we introduce some notation that will be used in
the paper:
\bi
\item $p_X(x)$ is the probability mass function of a discrete random variable $X$, or a probability density function of a continuous random variable $X$, and is simplified as $p(x)$ with no confuse on results.
\item Italic letters (e.g. $X$) denote scalars; and
bold letters $\xp$ and $\Xbf$ denote column vectors and matrices,
respectively.
\item $\Ibf$ denotes the identity matrix
and $\0bf$ denotes the all-zero vector or matrix. The dimensions of $\Ibf$ and $\0bf$ are determined by the context.
\item $|\Xbf|$, $\Xbf^T$, $\Xbf^{-1}$ and $\rank(\Xbf)$ denote
respectively the determinant, transpose,
inverse, and rank of the matrix $\Xbf$, $\|\xp\|$ denotes the Euclidean vector norm of $\xp$, i.e., $\|\xp\|^2=\xp^T\xp$, and $\otimes$ denotes the Kronecker product of matrices.
\item $\textrm{sign}(x)=1$ if $x\geq 0$ and $\textrm{sign}(x)=-1$ if $x<0$.
\item
$\xp^n=\left[\xp_1^T,\xp_2^T,\dots,\xp_n^T\right]^T$ is a long
vector that consists of a sequence of vectors $\xp_i, i=1,\dots,
n$.
\item $\diag[X_1,\cdots,X_n]$ is a diagonal matrix with diagonal entries $X_i$.
\item $\Vec\left(\Abf\right)$ denote the vectorization operator, i.e., let $\Abf=[\abf_1,\abf_2,\cdots,\abf_n]$, and $\abf_i,i=1,\cdots,n$ be the column vectors, then $\Vec\left(\Abf\right)=[\abf_1^T,\abf_2^T,\cdots,\abf_n^T]^T$.
\item $\theta=\textrm{atan}(x)$ means $\tan\theta=x$ and $\theta\in\left(-\frac{\pi}{2},\frac{\pi}{2}\right)$.
\item
$\xp\sim\Nmat\left(\0bf,\Sigmabf\right)$ means that the random
vector $\xp$ has the Gaussian distribution with zero mean and covariance
matrix $\Sigmabf$.
\item $E[\cdot]$ denotes expectation;
$\textrm{Cov}(\cdot)$ denotes covariance matrix; $I(\cdot;\cdot)$
denotes mutual information; $h(\cdot)$ denotes differential
entropy with the logarithmic base $e$, and
$\log(\cdot)=\log_e(\cdot)$.
\ei

\section{MIMO ICs}
In this section we derive sufficient conditions for a MIMO IC to have generally strong interference by comparing a special case of the Han and Kobayashi inner bound \cite{Han&Kobayashi:81IT} with a new outer bound.

\subsection{Inner bound}
We first obtain the achievable region by jointly decoding the signal and the interference.We also show that this region is a special case of Han and Kobayashi's achievable region despite the fact that it has a different expression from the Han and Kobayashi achievable region for the same coding scheme. Then, we apply this achievable region to MIMO ICs.
\begin{lemma}
The following rate region is achievable for a discrete memoryless IC
\bsq
0\leq R_1&{}\leq{}& I\left(X_1;Y_1|X_2Q\right)
\label{eq:innerR1}\\
0\leq R_2&{}\leq{}& I\left(X_2;Y_2|X_1Q\right)
\label{eq:innerR2}\\
R_1+R_2&{}\leq{}& I\left(X_1X_2;Y_1|Q\right)
\label{eq:innerRs1}\\
R_1+R_2&{}\leq{}& I\left(X_1X_2;Y_2|Q\right)
\label{eq:innerRs2}
\esq
where the input distribution factors as $ p\left(x_1x_2q\right)=p(q)p(x_1|q)p(x_2|q)$.
\label{lemma:inner}
\end{lemma}

The proof is given in Appendix \ref{appendix:inner} and is based on the analysis of error probability. In this proof, we require receiver $i$ to decode the message by searching the joint typical sequence set $A^{n}_\epsilon\left(QX_1X_2Y_i\right)$, $i=1,2$. We emphasize here that joint decoding means, e.g., receiver $1$ must correctly decode $X_1$ whereas the decoding for $X_2$ can be incorrect, i.e., its error probability is (\ref{eq:pe1}) instead of
\bqa
\textrm{Pr}\left\{{E_{11}^1}^c\bigcup\cup_{(i\neq 1,\textrm{any }j)}E_{ij}^1\bigcup\cup_{(j\neq 1,\textrm{any }i)}E_{ij}^1\right\}.
\eqa

If we consider the Han and Kobayashi achievable region in the simplified expression \cite{Han&Kobayashi:81IT,Chong-etal:08IT,Kramer:06Zurich}, then our coding scheme is equivalent to letting $W_1=X_1$ and $W_2=X_2$ in \cite[eqs. (11)-(18)]{Chong-etal:08IT}. However, it is interesting that by letting $W_1=X_1$ and $W_2=X_2$, \cite[eqs. (11)-(18)]{Chong-etal:08IT} become a region defined by (\ref{eq:innerR1})-(\ref{eq:innerRs2}) with an extra constraint:
\bqa
R_1+R_2\leq I\left(X_1;Y_2|X_2Q\right)+I\left(X_2;Y_1|X_1Q\right).
\label{eq:extra}
\eqa
This apparent inconsistency is caused by the fact that the rate constraint $S_1+T_2\leq I\left(W_2X_1;Y_1|W_1Q\right)$ in \cite[eq. (76)]{Chong-etal:08IT} is redundant when $W_1=X_1$ and $W_2=X_2$ (similarly, $S_2+T_1\leq I\left(W_1X_2;Y_2|W_2Q\right)$ is also redundant). This extra constraint (\ref{eq:extra}) is associated with receiver $i$'s error probability of decoding its own messages that are not carried by $W_i$. Therefore, when $W_i=X_i$, user $i$'s messages are all carried by $W_i$ and this extra constraint is redundant. Therefore, even if (\ref{eq:extra}) is violated, it does not contribute to the overall error probability of user $i$.

In fact the achievable region in Lemma \ref{lemma:inner} is still a subset of the Han and Kobayashi region. We state it formally in the following lemma, the proof of which is given in Appendix \ref{appendix:subset}.
\begin{lemma}
The achievable region in Lemma \ref{lemma:inner} is a subset of the Han and Kobayashi region.
\label{lemma:subset}
\end{lemma}

With Lemma \ref{lemma:inner}, we obtain the achievable rate region for a MIMO IC by jointly decoding the signal and the interference in the following lemma. We note that the time sharing procedure is unnecessary since all the constraints are concave functions.
\begin{lemma}
The following region is achievable for a MIMO IC:
\bqa
\bigcup_{\Sbf_i\succeq\0bf,\tr(\Sbf_i)\leq P_i,i=1,2}\left\{\begin{array}{l}
    0\leq R_1\leq g_1\left(\Sbf_1\right) \\
    0\leq R_2\leq g_2\left(\Sbf_2\right) \\
    R_1+R_2\leq g_{s1}\left(\Sbf_1,\Sbf_2\right)\\
    R_1+R_2\leq g_{s2}\left(\Sbf_1,\Sbf_2\right)
                                                    \end{array}
\right\}
\label{eq:MIMOhk}
\eqa
where
\bsq
g_1(\Sbf_1)&{}={}&\frac{1}{2}\log\left|\Ibf+\Hbf_1\Sbf_1\Hbf_1^T\right|
\label{eq:g1}\\
g_2(\Sbf_2)&{}={}&\frac{1}{2}\log\left|\Ibf+\Hbf_2\Sbf_2\Hbf_2^T\right|
\label{eq:g2}\\
g_{s1}(\Sbf_1,\Sbf_2)&{}={}&\frac{1}{2}\log\left|\Ibf+\Hbf_1\Sbf_1\Hbf_1^T+\Fbf_2\Sbf_2\Fbf_2^T\right|
\label{eq:gs1}\\
g_{s2}(\Sbf_1,\Sbf_2)&{}={}&\frac{1}{2}\log\left|\Ibf+\Hbf_2\Sbf_2\Hbf_2^T+\Fbf_1\Sbf_1\Fbf_1^T\right|.
\label{eq:gs2}
\esq
\label{lemma:MIMOhk}
\end{lemma}
We now proceed to obtain the maximum sum rate and other boundary points of region (\ref{eq:MIMOhk}).
\begin{lemma}
The maximum sum rate of (\ref{eq:MIMOhk}) is the maximum in the following optimization problem:
\bqa
\max &&\quad R_1+R_2\nn\\
\textrm{subject to} && \quad R_1+R_2\leq  g_1(\Sbf_1)+g_2(\Sbf_2)\nn\\
&&\quad R_1+R_2\leq g_{s1}(\Sbf_1,\Sbf_2)\nn\\
&&\quad R_1+R_2\leq g_{s2}(\Sbf_1,\Sbf_2)\nn\\
&&\quad \tr\left(\Sbf_i\right)\leq P_i,\quad \Sbf_i\succeq\0bf,\quad i=1,2.
\label{eq:MIMOsuml}
\eqa
Furthermore, if $\Sbf_i^*$, $i=1,2$ is optimal for problem (\ref{eq:MIMOsuml}), then there exist Lagrangian multipliers $\gamma,\lambda_i,\eta_i$ and $\Wbf_i$ that satisfy
\bsq
&&\gamma+\lambda_{1}+\lambda_{2}=1
\label{eq:KKTrateSl}\\
&&\Wbf_1=-\frac{\gamma}{2}\Hbf_1^T\left(\Ibf+\Hbf_1\Sbf_1^*\Hbf_1^T\right)^{-1}\Hbf_1-\frac{\lambda_{1}}{2}
\Hbf_1^T\left(\Ibf+\Hbf_1\Sbf_1^*\Hbf_1^T+\Fbf_2\Sbf_2^*\Fbf_2^T\right)^{-1}\Hbf_1\nn\\
    &&\hspace{.45in}-\frac{\lambda_{2}}{2}\Fbf_1^T\left(\Ibf+\Hbf_2\Sbf_2^*\Hbf_2^T+\Fbf_1\Sbf_1^*\Fbf_1^T\right)^{-1}\Fbf_1+\eta_1\Ibf
\label{eq:KKTW1}\\
&&\Wbf_2=-\frac{\gamma}{2}\Hbf_2^T\left(\Ibf+\Hbf_2\Sbf_2^*\Hbf_2^T\right)^{-1}\Hbf_2-\frac{\lambda_{1}}{2}
\Fbf_2^T\left(\Ibf+\Hbf_1\Sbf_1^*\Hbf_1^T+\Fbf_2\Sbf_2^*\Fbf_2^T\right)^{-1}\Fbf_2\nn\\
    &&\hspace{.45in}-\frac{\lambda_{2}}{2}\Hbf_2^T\left(\Ibf+\Hbf_2\Sbf_2^*\Hbf_2^T+\Fbf_1\Sbf_1^*\Fbf_1^T\right)^{-1}\Hbf_2+\eta_2\Ibf
\label{eq:KKTW2}\\
&&\gamma\left\{\begin{array}{cc}
                 >0&\qquad \textrm{if }R_1+R_2=g_1\left(\Sbf_1^*\right)+g_2\left(\Sbf_2^*\right) \\
                 =0&\qquad  \textrm{if }R_1+R_2<g_1\left(\Sbf_1^*\right)+g_2\left(\Sbf_2^*\right)
               \end{array}
\right.
\label{eq:gamma}\\
&&\lambda_{i}\left\{\begin{array}{cc}
                 >0&\qquad \textrm{if }R_1+R_2=g_{si}\left(\Sbf_1^*,\Sbf_2^*\right) \\
                 =0&\qquad  \textrm{if }R_1+R_2<g_{si}\left(\Sbf_1^*,\Sbf_2^*\right)
               \end{array}
\right.
\label{eq:KKTlambda}\\
&&\eta_i\left\{\begin{array}{cc}
                 >0&\qquad \textrm{if }\tr\left(\Sbf_i^*\right)=P_i \\
                 =0&\qquad  \textrm{if }\tr\left(\Sbf_i^*\right)<P_i
               \end{array}
\right.
\label{eq:KKTeta}\\
&&\tr\left(\Wbf_i\Sbf_i^*\right)=0
\label{eq:KKTWS}\\
&&\Wbf_i\succeq\0bf.
\label{eq:KKTW0}
\esq
\label{lemma:MIMOSl}
\end{lemma}
\bpf
Conditions (\ref{eq:KKTrateSl})-(\ref{eq:KKTW0}) are the Karush-Kuhn-Tucker (KKT) conditions of problem (\ref{eq:MIMOsuml}). The corresponding Lagrangian is
\bqa
L&{}={}&-(R_1+R_2)+\gamma\left(R_1+R_2-g_1-g_2\right)+\sum_{i=1}^2\lambda_i\left(R_1+R_2-g_{si}\right)+\sum_{i=1}^2\eta_i\left(\tr(\Sbf_i)-P_i\right)\nn\\
&&+\sum_{i=1}^2\tr\left(\Wbf_i\Sbf_i\right).
\eqa
Since (\ref{eq:MIMOsuml}) is a convex optimization problem, the Lagrangian multipliers do exist.
\epf
\begin{lemma}
The boundary points of the region defined in (\ref{eq:MIMOhk}) is determined by
\bqa
\bigcup_{0\leq r\leq\max\frac{1}{2}\log\left|\Ibf+\Hbf_2\Sbf_2\Hbf_2^T\right|}
\left\{R_1=R_1^*\left(r\right),\quad R_2=r \right\}
\eqa
where $R_1^*\left(r\right)$ is the maximum of the following optimization problem
\bqa
\max&&\quad R_1\nn\\
\textrm{subject to}&&\quad R_1\leq g_1\left(\Sbf_1\right)\nn\\
&&\quad r\leq g_2\left(\Sbf_2\right)\nn\\
&&\quad R_1\leq g_{s1}\left(\Sbf_1,\Sbf_2\right)-r\nn\\
&&\quad R_1\leq g_{s2}\left(\Sbf_1,\Sbf_2\right)-r\nn\\
&&\quad \tr\left(\Sbf_i\right)\leq P_i,\quad \Sbf_i\succeq\0bf.
\label{eq:MIMOIB}
\eqa
Furthermore, if $\Sbf_1^*$ and $\Sbf_2^*$ are optimal for problem (\ref{eq:MIMOIB}), then there exist Lagrangian multipliers $\alpha_i,\beta_i,\nu_i$ and $\Kbf_i$ that satisfy
\bsq
&&\alpha_1+\beta_1+\beta_2=1
\label{eq:KKTrateI}\\
&&\Kbf_1=-\frac{\alpha_1}{2}\Hbf_1^T\left(\Ibf+\Hbf_1\Sbf_1^*\Hbf_1^T\right)^{-1}\Hbf_1-\frac{\beta_1}{2}
\Hbf_1^T\left(\Ibf+\Hbf_1\Sbf_1^*\Hbf_1^T+\Fbf_2\Sbf_2^*\Fbf_2^T\right)^{-1}\Hbf_1\nn\\
    &&\hspace{.45in}-\frac{\beta_2}{2}\Fbf_1^T\left(\Ibf+\Hbf_2\Sbf_2^*\Hbf_2^T+\Fbf_1\Sbf_1^*\Fbf_1^T\right)^{-1}\Fbf_1+\nu_1\Ibf\\
&&\Kbf_2=-\frac{\alpha_2}{2}\Hbf_1^T\left(\Ibf+\Hbf_2\Sbf_2^*\Hbf_2^T\right)^{-1}\Hbf_2-\frac{\beta_1}{2}
\Fbf_2^T\left(\Ibf+\Hbf_1\Sbf_1^*\Hbf_1^T+\Fbf_2\Sbf_2^*\Fbf_2^T\right)^{-1}\Fbf_2\nn\\
    &&\hspace{.45in}-\frac{\beta_2}{2}\Hbf_2^T\left(\Ibf+\Hbf_2\Sbf_2^*\Hbf_2^T+\Fbf_1\Sbf_1^*\Fbf_1^T\right)^{-1}\Hbf_2+\nu_2\Ibf\\
&&\alpha_i\left\{\begin{array}{cc}
                 >0&\quad \textrm{if }R_i=g_i\left(\Sbf_i^*\right) \\
                 =0&\quad  \textrm{if }R_i<g_i\left(\Sbf_i^*\right)
               \end{array}
\right.\\
&&\beta_i\left\{\begin{array}{cc}
                 >0&\quad \textrm{if }R_1=g_{si}\left(\Sbf_1^*,\Sbf_2^*\right)-r \\
                 =0&\quad  \textrm{if }R_1<g_{si}\left(\Sbf_1^*,\Sbf_2^*\right)-r
               \end{array}
\right.\\
&&\nu_i\left\{\begin{array}{cc}
                 >0&\quad \textrm{if }\tr\left(\Sbf_i^*\right)=P_i \\
                 =0&\quad  \textrm{if }\tr\left(\Sbf_i^*\right)<P_i
               \end{array}
\right.\\
&&\tr\left(\Kbf_i\Sbf_i^*\right)=0\\
&&\Kbf_i\succeq\0bf.
\label{eq:KKTK0}
\esq
\label{lemma:MIMOIB}
\end{lemma}
\bpf
We first prove that $\{R_1=R_1^*(r),R_2=r\}$ is a boundary point of the region given in (\ref{eq:MIMOhk}). By the constraint conditions of (\ref{eq:MIMOIB}), the rate pair $\{R_1^*(r),r\}$ belongs to the region (\ref{eq:MIMOhk}) determined by $\Sbf_1^*$ and $\Sbf_2^*$. Therefore, $\{R_1^*(r),r\}$ is in the set (\ref{eq:MIMOhk}). Next we assume, on the contrary, that $\{R_1^*(r),r\}$ is not on the boundary. Then there exists a rate pair $\{R_1^\prime,r\}$ with $R_1^\prime>R_1^*(r)$ which is also in region (\ref{eq:MIMOhk}). Therefore, there exist matrices $\Sbf_i^\prime$ with $\tr\left(\Sbf_i^\prime\right)\leq P_i$ and $\Sbf_i^\prime\succeq\0bf$, $i=1,2$,  such that
\bqa
R_1^\prime&{}\leq{}& g_1\left(\Sbf_1^\prime\right)\nn\\
r&{}\leq{}& g_2\left(\Sbf_2^\prime\right)\nn\\
R_1^\prime+r&{}\leq{}& g_{s1}\left(\Sbf_1^\prime,\Sbf_2^\prime\right)\nn\\
R_1^\prime+r&{}\leq{}& g_{ss}\left(\Sbf_1^\prime,\Sbf_2^\prime\right)\nn.
\eqa
Thus, $\{R_1^\prime,r\}$ is feasible for optimization problem (\ref{eq:MIMOIB}) and hence $R_1^\prime\leq R_1^*(r)$ since $R_1^*(r)$ is the maximum of problem (\ref{eq:MIMOIB}). This contradicts our assumption  $R_1^\prime>R_1^*(r)$. Therefore, $\{R_1^*(r),r\}$ is on the boundary.

Conditions (\ref{eq:KKTrateI})-(\ref{eq:KKTK0}) are the KKT conditions of problem (\ref{eq:MIMOIB}). The corresponding Lagrangian is
\bqa
L&{}={}&-R_1+\alpha_1\left(R_1-g_1\right)+\alpha_2\left(r-g_2\right)+\sum_{i=1}^2\beta_i\left(R_1-g_{si}+r\right)
    +\sum_{i=1}^2\nu_i\left(\tr(\Sbf_i)-P_i\right)\nn\\
&&+\sum_{i=1}^2\tr\left(\Kbf_i\Sbf_i\right).
\eqa
Since (\ref{eq:MIMOIB}) is a convex optimization problem, the Lagrangian multipliers do exist.
\epf

\subsection{Outer bounds}
The outer bound is obtained by providing additional information to both receivers.
\begin{lemma}
The closure of the following set\footnote{Obviously, any of the constraints in (\ref{eq:MIMOo}) can be removed, and the closure of the resulting set is still an outer bound on the capacity region.} is an outer bound on the capacity region of a MIMO IC:
\bqa
\bigcup_{\Sbf_i\succeq\0bf,\tr(\Sbf_i)\leq P_i,i=1,2}\left\{\begin{array}{l}
    R_1\leq g_1(\Sbf_1) \\
    R_2\leq g_2(\Sbf_2) \\
    R_1+R_2\leq \bar g_{s1}(\Sbf_1,\Sbf_2)\\
    R_1+R_2\leq \bar g_{s2}(\Sbf_1,\Sbf_2)
                                                            \end{array}
\right\}
\label{eq:MIMOo}
\eqa
where $g_1$ and $g_2$ are defined in (\ref{eq:g1}) and (\ref{eq:g2}), respectively, and
\bqa
\bar g_{s1}(\Sbf_1,\Sbf_2)&{}={}&\frac{1}{2}\log\left|\Ibf+\Hbf_1\Sbf_1\Hbf_1^T\left(\Ibf+\Fbf_2\Sbf_2\Fbf_2^T\right)^{-1}\right|
+\frac{1}{2}\log\left|\Ibf+\left[\begin{array}{c}
                                                    \Hbf_2 \\
                                                    \Fbf_2
                                                  \end{array}
\right]\Sbf_2\left[\begin{array}{c}
                                                    \Hbf_2 \\
                                                    \Fbf_2
                                                  \end{array}
\right]^T\Ebf_2^{-1}\right|\\
    &{}={}&\frac{1}{2}\log\left|\Ibf+\Hbf_1\Sbf_1\Hbf_1^T\left(\Ibf+\Fbf_2\Sbf_2\Fbf_2^T\right)^{-1}\right|+\frac{1}{2}\log\left|\Ibf+\Sbf_2
    \Fbf_2^T\Fbf_2+2\Sbf_2\Obf_2\right|
\label{eq:gs1bar}\\
\bar g_{s2}(\Sbf_1,\Sbf_2)&{}={}&\frac{1}{2}\log\left|\Ibf+\Hbf_2\Sbf_2\Hbf_2^T\left(\Ibf+\Fbf_1\Sbf_1\Fbf_1^T\right)^{-1}\right|
+\frac{1}{2}\log\left|\Ibf+\left[\begin{array}{c}
                                                    \Hbf_1 \\
                                                    \Fbf_1
                                                  \end{array}
\right]\Sbf_1\left[\begin{array}{c}
                                                    \Hbf_1 \\
                                                    \Fbf_1
                                                  \end{array}
\right]^T\Ebf_1^{-1}\right|\\
    &{}={}&\frac{1}{2}\log\left|\Ibf+\Hbf_2\Sbf_2\Hbf_2^T\left(\Ibf+\Fbf_1\Sbf_1\Fbf_1^T\right)^{-1}\right|+\frac{1}{2}\log\left|\Ibf+\Sbf_1
    \Fbf_1^T\Fbf_1+2\Sbf_1\Obf_1\right|
\label{eq:gs2bar}
\eqa
and $\Ebf_i$ and $\Obf_i$, $i=1,2$, are defined as\footnote{We note that the $\Ibf$ of (\ref{eq:Ei}) in the first row has dimension $r_i$ and the $\Ibf$ in the second row has dimension $r_j$ where $j\in\{1,2\},j\neq i$.}
\bqa
\Ebf_i&{}={}&\left[\begin{array}{cc}
               \Ibf &\quad \Abf_i \\
               \Abf_i^T &\quad \Ibf
             \end{array}
\right]\succ\0bf.
\label{eq:Ei}\\
\Obf_i&{}={}&\frac{1}{2}\left(\Hbf_i-\Abf_i\Fbf_i\right)^T\left(\Ibf-\Abf_i\Abf_i^T\right)^{-1}\left(\Hbf_i-\Abf_i\Fbf_i\right).
\label{eq:O}
\eqa
\label{lemma:MIMOo}
\end{lemma}
\bpf
Let $\xp_i^n=\left[\xp_{i1}^T,\cdots,\xp_{in}^T\right]^T$ be an input sequence of user $i$ that satisfies
\bqa
\sum_{j=1}^n\Cov\left(\xp_{ij}\right)&{}={}&n\Sbf_i
\label{eq:pcons1}\\
\tr\left(\Sbf_i\right)&{}\leq{}&P_i.
\label{eq:pcons2}
\eqa
Then we immediately obtain the $R_1\leq g_1\left(\Sbf_1\right)$ and $R_2\leq g_2\left(\Sbf_2\right)$ in (\ref{eq:MIMOo}). For $\epsilon>0$ and $\epsilon\rightarrow 0$ when $n\rightarrow\infty$, by Fano's inequality we have
\bqa
&&n(R_1+R_2)-n\epsilon\nn\\
&&\leq I\left(\xp_1^n;\yp_1^n\right)+I\left(\xp_2^n;\yp_2^n\right)\nn\\
&&\stackrel{(a)}\leq I\left(\xp_1^n;\yp_1^n\right)+I\left(\xp_2^n;\yp_2^n,\xp_1^n,\Fbf_2\xp_2^n+\np_2^n\right)\nn\\
&&=h\left(\Hbf_1\xp_1^n+\Fbf_2\xp_2^n+\zp_1^n\right)-h\left(\Fbf_2\xp_2^n+\zp_1^n\right)+h\left(\Fbf_2\xp_2^n+\np_2^n\right)-h\left(\np_2^n\right)
    +h\left(\Fbf_2\xp_2^n+\zp_2^n|\Fbf_2\xp_2^n+\np_2^n\right)\nn\\
 &&\hspace{.2in}-h\left(\zp_2^n|\np_2^n\right)\nn\\
&&\stackrel{(b)}=h\left(\Hbf_1\xp_1^n+\Fbf_2\xp_2^n+\zp_1^n\right)-nh\left(\Fbf_2\xp_{2G}+\zp_1\right)+nh\left(\Fbf_2\xp_{2G}+\np_2\right)
-h\left(\np_2^n\right)
    +h\left(\Fbf_2\xp_2^n+\zp_2^n|\Fbf_2\xp_2^n+\np_2^n\right)\nn\\
    &&\hspace{.2in}-h\left(\zp_2^n|\np_2^n\right)\nn\\
&&\stackrel{(c)}\leq nh\left(\Hbf_1\xp_{1G}+\Fbf_2\xp_{2G}+\zp_1\right)-nh\left(\Fbf_2\xp_{2G}+\zp_1\right)+nh\left(\Fbf_2\xp_{2G}+\np_2\right)-nh\left(\np_2\right)\nn\\
    &&\hspace{.2in}+nh\left(\Fbf_2\xp_{2G}+\zp_2|\Fbf_2\xp_{2G}+\np_2\right)-nh\left(\zp_2|\np_2\right)\nn\\
&&=nI\left(\xp_{1G};\Hbf_1\xp_{1G}+\Fbf_2\xp_{2G}+\zp_1\right)+nI\left(\xp_{2G};\left[\begin{array}{c}
                                                                                        \Hbf_2 \\
                                                                                        \Fbf_2
                                                                                      \end{array}
\right]\xp_{2G}+\left[\begin{array}{c}
                                                                                        \zp_2 \\
                                                                                        \np_2
                                                                                      \end{array}
\right]\right)\nn\\
&&=\frac{1}{2}\log\left|\Ibf+\Hbf_1\Sbf_1\Hbf_1^T\left(\Ibf+\Fbf_2\Sbf_2\Fbf_2^T\right)^{-1}\right|+\frac{1}{2}\log\left|\Ibf+\left[\begin{array}{c}
                                                    \Hbf_2 \\
                                                    \Fbf_2
                                                  \end{array}
\right]\Sbf_2\left[\begin{array}{c}
                                                    \Hbf_2 \\
                                                    \Fbf_2
                                                  \end{array}
\right]^T\Ebf_2^{-1}\right|
\eqa
where, in (a) we let $\np_2^n$ be a sequence of independent and identically distributed Gaussian vectors each has the following joint distribution with $\zp_2$:
\bqa
\left[\begin{array}{c}
        \zp_2 \\
        \np_2
      \end{array}
\right]\sim\Nmat\left(\0bf,\Ebf_2\right)=\Nmat\left(\0bf,\left[\begin{array}{cc}
               \Ibf &\quad \Abf_2 \\
               \Abf_2^T &\quad \Ibf
             \end{array}
\right]\right).
\eqa
Equality (b) is by the fact that $\np_2$ and $\zp_1$ have identical marginal distributions, and thus
\bqa
&&-h\left(\Fbf_2\xp_2^n+\zp_1^n\right)+h\left(\Fbf_2\xp_2^n+\np_2^n\right)\nn\\
&&=0\nn\\
&&=-nh\left(\Fbf_2\xp_{2G}+\zp_1\right)+nh\left(\Fbf_2\xp_{2G}+\np_2\right)\nn
\eqa
where
\bqa
\xp_{iG}\sim\Nmat\left(\0bf,\Sbf_i\right).\nn
\eqa
Inequality (c) is by \cite[Lemma 2]{Shang-etal:10IT_mimo}.

To show (\ref{eq:gs1bar}), we have
\bqa
&&\log\left|\Ibf+\left[\begin{array}{c}
                                                    \Hbf_2 \\
                                                    \Fbf_2
                                                  \end{array}
\right]\Sbf_2\left[\begin{array}{c}
                                                    \Hbf_2 \\
                                                    \Fbf_2
                                                  \end{array}
\right]^T\Ebf_2^{-1}\right|\nn\\
&&\stackrel{(a)}=\log\left|\Ibf+\Sbf_2\left[\begin{array}{c}
                                                    \Hbf_2 \\
                                                    \Fbf_2
                                                  \end{array}
\right]^T\left[\begin{array}{cc}
                 \Ibf & \quad \Abf_2 \\
                 \Abf_2^T & \Ibf
               \end{array}
\right]^{-1}\left[\begin{array}{c}
                                                    \Hbf_2 \\
                                                    \Fbf_2
                                                  \end{array}
\right]\right|\nn\\
&&=\log\left|\Ibf+\Sbf_2\left[\begin{array}{c}
                                                    \Fbf_2 \\
                                                    \Hbf_2
                                                  \end{array}
\right]^T\left[\begin{array}{cc}
                 \Ibf & \quad \Abf_2^T \\
                 \Abf_2 & \Ibf
               \end{array}
\right]^{-1}\left[\begin{array}{c}
                                                    \Fbf_2 \\
                                                    \Hbf_2
                                                  \end{array}
\right]\right|\nn\\
&&\stackrel{(b)}=\log\left|\Ibf+\Sbf_2\left[\begin{array}{c}
                                                    \Fbf_2 \\
                                                    \Hbf_2
                                                  \end{array}
\right]^T\left(\left[\begin{array}{cc}
                 \Ibf & \quad \0bf \\
                 \0bf & \0bf
               \end{array}
\right]+\left[\begin{array}{c}
                \Abf_2^T \\
                -\Ibf
              \end{array}
\right]\left(\Ibf-\Abf_2\Abf_2^T\right)^{-1}\left[\Abf_2,\quad-\Ibf\right]\right)\left[\begin{array}{c}
                                                    \Fbf_2 \\
                                                    \Hbf_2
                                                  \end{array}
\right]\right|\nn\\
&&=\log\left|\Ibf+\Sbf_2
    \Fbf_2\Fbf_2^T+\Sbf_2\left(\Hbf_2-\Abf_2\Fbf_2\right)^T\left(\Ibf-\Abf_2\Abf_2^T\right)^{-1}\left(\Hbf_2-\Abf_2\Fbf_2\right)\right|\nn\\
&&=\log\left|\Ibf+\Sbf_2
    \Fbf_2\Fbf_2^T+2\Sbf_2\Obf_2\right|
\eqa
where (a) is by the matrix identity
\bqa
\left|\Ibf+\Abf\Bbf\right|=\left|\Ibf+\Bbf\Abf\right|
\eqa
and (b) is by \cite[Lemma 3]{Shang&Poor:11IT_submission_MIMO}. The other sum-rate bound $\bar g_{s2}$ is similarly obtained.

We have established the fact that for any input sequences $\xp_1^n$ and $\xp_2^n$ that satisfy (\ref{eq:pcons1}) and (\ref{eq:pcons2}), the corresponding rate pair is bounded by
\bqa
R_i&{}\leq{}&g_i(\Sbf_i)\\
R_1+R_2&{}\leq{}& \bar g_{si}(\Sbf_1,\Sbf_2).
\eqa
Therefore, (\ref{eq:MIMOo}) is an outer bound for the capacity region.
\epf
\begin{lemma}
The $\bar g_{s1}$ and $\bar g_{s2}$ are both concave functions of $\Sbf_1$ and $\Sbf_2$ for any $\Ebf_1$ and $\Ebf_2$ that satisfy (\ref{eq:Ei}).
\label{lemma:convex}
\end{lemma}
\bpf
This is an immediate result of \cite[Lemma 2]{Shang&Poor:11IT_submission_MIMO}. Considering \cite[eq.(16)]{Shang&Poor:11IT_submission_MIMO}, if we choose $\Ebf_1=\Ibf$ and $\Ebf_2$ as in (\ref{eq:Ei}), then \cite[eq.(16)]{Shang&Poor:11IT_submission_MIMO} reduces to $\bar g_{s1}$. Similarly, if we choose $\Ebf_2=\Ibf$ and $\Ebf_1$ as in (\ref{eq:Ei}), then \cite[eq.(16)]{Shang&Poor:11IT_submission_MIMO} reduces to $\bar g_{s2}$. Therefore, $\bar g_{s1}$ and $\bar g_{s2}$ are both concave functions.
\epf

Using Lemmas \ref{lemma:MIMOo} and \ref{lemma:convex}, we obtain the maximal sum-rate and the boundaries of the outer bound in the following lemmas.
\begin{lemma}
The maximum in the following optimization problem is an upper bound on the sum-rate capacity of the MIMO IC:
\bqa
\max &&\quad R_1+R_2\nn\\
\textrm{subject to} && \quad R_1+R_2\leq  g_1(\Sbf_1)+g_2(\Sbf_2)\nn\\
&&\quad R_1+R_2\leq \bar g_{s1}(\Sbf_1,\Sbf_2)\nn\\
&&\quad R_1+R_2\leq \bar g_{s2}(\Sbf_1,\Sbf_2)\nn\\
&&\quad \tr\left(\Sbf_i\right)\leq P_i,\quad \Sbf_i\succeq\0bf,\quad i=1,2.
\label{eq:MIMOsumu}
\eqa
Furthermore, if $\Sbf_1^*$ and $\Sbf_2^*$ are optimal for problem (\ref{eq:MIMOsumu}), and there exist matrices $\Abf_i$, $i=1,2$, that satisfy
\bqa
\Sbf_i^*\Hbf_i^T&{}={}&\Sbf_i^*\Fbf_i^T\Abf_i^T
\label{eq:Markov}\\
\Abf_i\Abf_i^T&{}\preceq{}&\Ibf
\label{eq:A}
\eqa
for $i=1,2$, then there exist Lagrangian multipliers $\bar\gamma,\bar\lambda_i,\bar\eta_i$ and $\overline\Wbf_i$ that satisfy
\bsq
&&\bar\gamma+\bar\lambda_{1}+\bar\lambda_{2}=1
\label{eq:KKTrateSu}\\
&&\overline\Wbf_1=-\frac{\bar\gamma}{2}\Hbf_1^T\left(\Ibf+\Hbf_1\Sbf_1^*\Hbf_1^T\right)^{-1}\Hbf_1-\frac{\bar\lambda_{1}}{2}
\Hbf_1^T\left(\Ibf+\Hbf_1\Sbf_1^*\Hbf_1^T+\Fbf_2\Sbf_2^*\Fbf_2^T\right)^{-1}\Hbf_1\nn\\
    &&\hspace{.45in}-\frac{\bar\lambda_{2}}{2}\Fbf_1^T\left(\Ibf+\Hbf_2\Sbf_2^*\Hbf_2^T+\Fbf_1\Sbf_1^*\Fbf_1^T\right)^{-1}\Fbf_1+\bar\eta_1\Ibf
    -\bar\lambda_2\Obf_1\\
&&\overline\Wbf_2=-\frac{\bar\gamma}{2}\Hbf_2^T\left(\Ibf+\Hbf_2\Sbf_2^*\Hbf_2^T\right)^{-1}\Hbf_2-\frac{\bar\lambda_{1}}{2}
\Fbf_2^T\left(\Ibf+\Hbf_1\Sbf_1^*\Hbf_1^T+\Fbf_2\Sbf_2^*\Fbf_2^T\right)^{-1}\Fbf_2\nn\\
    &&\hspace{.45in}-\frac{\bar\lambda_{2}}{2}\Hbf_2^T\left(\Ibf+\Hbf_2\Sbf_2^*\Hbf_2^T+\Fbf_1\Sbf_1^*\Fbf_1^T\right)^{-1}\Hbf_2+\bar\eta_2\Ibf
    -\bar\lambda_1\Obf_2\\
&&\bar\gamma\left\{\begin{array}{cc}
                 >0&\quad \textrm{if }R_1+R_2=g_1\left(\Sbf_1^*\right)+g_2\left(\Sbf_2^*\right) \\
                 =0&\quad  \textrm{if }R_1+R_2<g_1\left(\Sbf_1^*\right)+g_2\left(\Sbf_2^*\right)
               \end{array}
\right.\\
&&\bar\lambda_{i}\left\{\begin{array}{cc}
                 >0&\quad \textrm{if }R_1+R_2=\bar g_{si}\left(\Sbf_1^*,\Sbf_2^*\right)=g_{si}\left(\Sbf_1^*,\Sbf_2^*\right) \\
                 =0&\quad  \textrm{if }R_1+R_2<\bar g_{si}\left(\Sbf_1^*,\Sbf_2^*\right)=g_{si}\left(\Sbf_1^*,\Sbf_2^*\right)
               \end{array}
\right.\\
&&\bar\eta_i\left\{\begin{array}{cc}
                 >0&\quad \textrm{if }\tr\left(\Sbf_i^*\right)=P_i \\
                 =0&\quad  \textrm{if }\tr\left(\Sbf_i^*\right)<P_i
               \end{array}
\right.\\
&&\tr\left(\overline\Wbf_i\Sbf_i^*\right)=0\\
&&\overline\Wbf_i\succeq\0bf
\label{eq:KKTWSu}
\esq
for $i=1,2$, where $\Obf_i$ is defined in (\ref{eq:O}).
\label{lemma:MIMOSu}
\end{lemma}
\bpf
By Lemma \ref{lemma:convex}, (\ref{eq:MIMOsumu}) is a convex optimization problem; therefore, there exist Lagrangian multipliers that satisfy the KKT conditions (\ref{eq:KKTrateSu})-(\ref{eq:KKTWSu}). The corresponding Lagrangian is
\bqa
L&{}={}&-(R_1+R_2)+\bar\gamma\left(R_1+R_2-g_1-g_2\right)+\sum_{i=1}^2\bar\lambda_i\left(R_1+R_2-g_{si}\right)
    +\sum_{i=1}^2\bar\eta_i\left(\tr(\Sbf_i)-P_i\right)\nn\\
&&+\sum_{i=1}^2\tr\left(\overline\Wbf_i\Sbf_i\right).
\eqa
Thus, comparing to Lemma \ref{lemma:MIMOSl} we need only to show that for $i,j\in\{1,2\}$ and $i\neq j$,
\bqa
\bar g_{si}\left(\Sbf_1^*,\Sbf_2^*\right)&{}={}&g_{si}\left(\Sbf_1^*,\Sbf_2^*\right)
\label{eq:eqS}\\
\left.\frac{\partial \bar g_{si}}{\partial \Sbf_i}\right|_{\tiny\begin{array}{c}
                                                             \Sbf_1=\Sbf_1^* \\
                                                             \Sbf_2=\Sbf_2^*
                                                           \end{array}
}&{}={}&\left.\frac{\partial g_{si}}{\partial \Sbf_i}\right|_{\tiny\begin{array}{c}
                                                             \Sbf_1=\Sbf_1^* \\
                                                             \Sbf_2=\Sbf_2^*
                                                           \end{array}
}
\label{eq:partialSame}\\
\left.\frac{\partial \bar g_{si}}{\partial \Sbf_j}\right|_{\tiny\begin{array}{c}
                                                             \Sbf_1=\Sbf_1^* \\
                                                             \Sbf_2=\Sbf_2^*
                                                           \end{array}
}&{}={}&\left.\frac{\partial g_{si}}{\partial \Sbf_j}\right|_{\tiny\begin{array}{c}
                                                             \Sbf_1=\Sbf_1^* \\
                                                             \Sbf_2=\Sbf_2^*
                                                           \end{array}
}+\Obf_j.
\label{eq:partialDiff}
\eqa
Equalities (\ref{eq:eqS}) and (\ref{eq:partialSame}) are straightforward by (\ref{eq:Markov}). By symmetry, it suffices to show (\ref{eq:partialDiff}) for $i=1$ and $j=2$:
\bqa
&&\left.\frac{\partial \bar g_{s1}}{\partial \Sbf_2}\right|_{\tiny\begin{array}{c}
                                                             \Sbf_1=\Sbf_1^* \\
                                                             \Sbf_2=\Sbf_2^*
                                                           \end{array}}\nn\\
&&=\frac{1}{2}\Fbf_2^T\left(\Ibf+\Hbf_1\Sbf_1^*\Hbf_1^T+\Fbf_2\Sbf_2^*\Fbf_2^T\right)^{-1}\Fbf_2
-\frac{1}{2}\Fbf_2^T\left(\Ibf+\Fbf_2\Sbf_2^*\Fbf_2^T\right)^{-1}\Fbf_2\nn\\
    &&\hspace{.2in}+\frac{1}{2}\left(\Fbf_2^T\Fbf_2+2\Obf_2\right)\left(\Ibf+\Sbf_2^*
    \Fbf_2^T\Fbf_2+2\Sbf_2^*\Obf_2\right)^{-1}\nn\\
&&\stackrel{(a)}=\frac{1}{2}\Fbf_2^T\left(\Ibf+\Hbf_1\Sbf_1^*\Hbf_1^T+\Fbf_2\Sbf_2^*\Fbf_2^T\right)^{-1}\Fbf_2
-\frac{1}{2}\Fbf_2^T\left(\Ibf+\Fbf_2\Sbf_2^*\Fbf_2^T\right)^{-1}\Fbf_2\nn\\
    &&\hspace{.2in}+\frac{1}{2}\left(\Fbf_2^T\Fbf_2+2\Obf_2\right)\left(\Ibf+\Sbf_2^*\Fbf_2^T\Fbf_2\right)^{-1}\nn\\
&&\stackrel{(b)}=\frac{1}{2}\Fbf_2^T\left(\Ibf+\Hbf_1\Sbf_1^*\Hbf_1^T+\Fbf_2\Sbf_2^*\Fbf_2^T\right)^{-1}\Fbf_2
    +\Obf_2\left(\Ibf+\Sbf_2^*\Fbf_2^T\Fbf_2\right)^{-1}\nn\\
&&\stackrel{(c)}=\frac{1}{2}\Fbf_2^T\left(\Ibf+\Hbf_1\Sbf_1^*\Hbf_1^T+\Fbf_2\Sbf_2^*\Fbf_2^T\right)^{-1}\Fbf_2
    +\Obf_2\left(\Ibf-\Sbf_2^*\left(\Ibf+\Fbf_2^T\Fbf_2\Sbf_2^*\right)^{-1}\right)\Fbf_2^T\Fbf_2\nn\\
&&\stackrel{(d)}=\frac{1}{2}\Fbf_2^T\left(\Ibf+\Hbf_1\Sbf_1^*\Hbf_1^T+\Fbf_2\Sbf_2^*\Fbf_2^T\right)^{-1}\Fbf_2
    +\Obf_2\nn\\
&&=\left.\frac{\partial g_{s1}}{\partial \Sbf_2}\right|_{\tiny\begin{array}{c}
                                                             \Sbf_1=\Sbf_1^* \\
                                                             \Sbf_2=\Sbf_2^*
                                                           \end{array}}
    +\Obf_2
\eqa
where (a) and (d) are both from (\ref{eq:Markov}) which implies
\bqa
\Sbf_i^*\Obf_i^*=\0bf.
\eqa
Equality (b) is by the matrix identity \cite[p. 151]{Searle:book}:
\bqa
\Cbf\left(\Ibf+\Dbf\Cbf\right)^{-1}=\left(\Ibf+\Cbf\Dbf\right)^{-1}\Cbf
\label{eq:Searle1}
\eqa
which implies
\bqa
-\Fbf_2^T\left(\Ibf+\Fbf_2\Sbf_2^*\Fbf_2^T\right)^{-1}\Fbf_2+\Fbf_2^T\Fbf_2\left(\Ibf+\Sbf_2^*\Fbf_2^T\Fbf_2\right)^{-1}=\0bf;\nn
\eqa
and (c) is by the Woodbury matrix identity \cite[p. 19]{Horn&Johnson:book}:
\bqa
\left(\Cbf+\Ubf\Bbf\Vbf\right)^{-1}=\Cbf^{-1}-\Cbf^{-1}\Ubf\left(\Bbf^{-1}+\Vbf\Cbf^{-1}\Ubf\right)^{-1}\Vbf\Cbf^{-1}.
\label{eq:woodbury}
\eqa
\epf
\begin{lemma}
Let $R_2=r$ with $0\leq r\leq \max\frac{1}{2}\log\left|\Ibf+\Hbf_2\Sbf_2\Hbf_2\right|$, and let $\bar R_1^*(r)$ be the maximum in the following optimization problem:
\bqa
\max &&\quad R_1\nn\\
\textrm{subject to} && \quad R_1\leq  g_1(\Sbf_1)\nn\\
&&\quad r\leq g_2(\Sbf_2)\nn\\
&&\quad R_1\leq \bar g_{s1}(\Sbf_1,\Sbf_2)-r\nn\\
&&\quad R_1\leq \bar g_{s2}(\Sbf_1,\Sbf_2)-r\nn\\
&&\quad \tr\left(\Sbf_i\right)\leq P_i,\quad \Sbf_i\succeq\0bf,\quad i=1,2.
\label{eq:MIMOoB}
\eqa
Then $\left\{\bar R_1^*(r),r\right\}$ is on the boundary of the outer bound given in (\ref{eq:MIMOo}). Furthermore, if $\Sbf_1^*$ and $\Sbf_2^*$ are optimal for problem (\ref{eq:MIMOoB}), and there exist matrices $\Abf_i$, $i=1,2$, that satisfy (\ref{eq:Markov}) and (\ref{eq:A}), then there exist Lagrangian multipliers $\bar\alpha_i,\bar\beta_i,\bar\nu_i$ and $\overline\Kbf_i$ that satisfy
\bsq
&&\bar\alpha_1+\bar\beta_1+\bar\beta_2=1
\label{eq:KKTrateO}\\
&&\overline\Kbf_1=-\frac{\bar\alpha_1}{2}\Hbf_1^T\left(\Ibf+\Hbf_1\Sbf_1^*\Hbf_1^T\right)^{-1}\Hbf_1-\frac{\bar\beta_1}{2}
\Hbf_1^T\left(\Ibf+\Hbf_1\Sbf_1^*\Hbf_1^T+\Fbf_2\Sbf_2^*\Fbf_2^T\right)^{-1}\Hbf_1\nn\\
    &&\hspace{.45in}-\frac{\bar\beta_2}{2}\Fbf_1^T\left(\Ibf+\Hbf_2\Sbf_2^*\Hbf_2^T+\Fbf_1\Sbf_1^*\Fbf_1^T\right)^{-1}\Fbf_1+\bar\nu_1\Ibf
    -\bar\beta_2\Obf_1\\
&&\overline\Kbf_2=-\frac{\bar\alpha_2}{2}\Hbf_2^T\left(\Ibf+\Hbf_2\Sbf_2^*\Hbf_2^T\right)^{-1}\Hbf_2-\frac{\bar\beta_1}{2}
\Fbf_2^T\left(\Ibf+\Hbf_1\Sbf_1^*\Hbf_1^T+\Fbf_2\Sbf_2^*\Fbf_2^T\right)^{-1}\Fbf_2\nn\\
    &&\hspace{.45in}-\frac{\bar\beta_2}{2}\Hbf_2^T\left(\Ibf+\Hbf_2\Sbf_2^*\Hbf_2^T+\Fbf_1\Sbf_1^*\Fbf_1^T\right)^{-1}\Hbf_2+\bar\nu_2\Ibf
    -\bar\beta_1\Obf_2\\
&&\bar\alpha_i\left\{\begin{array}{cc}
                 >0&\quad \textrm{if }R_1=g_1\left(\Sbf_i^*\right) \\
                 =0&\quad  \textrm{if }R_1<g_1\left(\Sbf_i^*\right)
               \end{array}
\right.\\
&&\bar\beta_{i}\left\{\begin{array}{cc}
                 >0&\quad \textrm{if }R_1=\bar g_{si}\left(\Sbf_1^*,\Sbf_2^*\right)-r=g_{si}\left(\Sbf_1^*,\Sbf_2^*\right)-r \\
                 =0&\quad  \textrm{if }R_1<\bar g_{si}\left(\Sbf_1^*,\Sbf_2^*\right)-r=g_{si}\left(\Sbf_1^*,\Sbf_2^*\right)-r
               \end{array}
\right.\\
&&\bar\nu_i\left\{\begin{array}{cc}
                 >0&\quad \textrm{if }\tr\left(\Sbf_i^*\right)=P_i \\
                 =0&\quad  \textrm{if }\tr\left(\Sbf_i^*\right)<P_i
               \end{array}
\right.\\
&&\tr\left(\overline\Kbf_i\Sbf_i^*\right)=0\\
&&\overline\Kbf_i\succeq\0bf
\label{eq:KKTOK0}
\esq
for $i=1,2$, where $\Obf_i$ is defined in (\ref{eq:O}).
\label{lemma:MIMOoB}
\end{lemma}
\bpf
Similarly to the proof of Lemma \ref{lemma:MIMOIB}, it can be shown that $\left\{\bar R_1^*(r),r\right\}$ is on the boundary of the outer bound (\ref{eq:MIMOo}).
Conditions (\ref{eq:KKTrateO})-(\ref{eq:KKTOK0}) are the KKT conditions of problem (\ref{eq:MIMOoB}). The corresponding Lagrangian is
\bqa
L&{}={}&-R_1+\bar\alpha_1\left(R_1-g_1(\Sbf_1)\right)+\bar\alpha_2\left(r-g_2(\Sbf_2)\right)+\sum_{i=1}^2\bar\beta_i\left(R_1+r-\bar g_{si}\left(\Sbf_1,\Sbf_2\right)\right)+\sum_{i=1}^2\bar\nu_i\left(\tr(\Sbf_i)-P_i\right)\nn\\
&&+\sum_{i=1}^2\tr\left(\overline\Kbf_i\Sbf_i\right).
\eqa
Since (\ref{eq:MIMOIB}) is a convex optimization problem, the Lagrangian multipliers do exist. The rest of the proof is similar to that of Lemma \ref{lemma:MIMOSu} and is hence omitted.
\epf

\subsection{Sum-rate capacity and capacity region}
Now we obtain the capacity results for MIMO ICs with generally strong interference by comparing the inner and outer bounds.
\begin{theorem}
Suppose $\Sbf_i^*$ $i=1,2$, are maximizers of problem (\ref{eq:MIMOsuml}) and for $i=1,2$, let $\lambda_i$ and $\Wbf_i$ be the Lagrangian multipliers in (\ref{eq:KKTrateSl})-(\ref{eq:KKTW0}). For any $\lambda_j>0$, $j=1,2$, if there exist $\Abf_i$, $i=1,2$, $i\neq j$, that satisfy (\ref{eq:Markov}) and (\ref{eq:A}), and
\bqa
\Wbf_i\succeq \lambda_j\Obf_i
\eqa
where $\Obf_j$ is defined in (\ref{eq:O}), then the sum-rate capacity of the MIMO IC is the maximum in problem (\ref{eq:MIMOsuml}) and is achieved by the input distributions $\xp_i\sim\Nmat\left(\0bf,\Sbf_i^*\right)$, $i=1,2$, and jointly decoding the signal and the interference.
\label{theorem:MIMOs}
\end{theorem}
\bpf
Since $\Sbf_1^*$ and $\Sbf_2^*$ maximize problem (\ref{eq:MIMOsuml}), the KKT conditions in (\ref{eq:KKTrateSl})-(\ref{eq:KKTW0}) hold.

If $\lambda_1=\lambda_2=0$, the maximal achievable sum rate is $\max_{\tr(\Sbf_i)\leq P_i,\Sbf_i\succeq\0bf} \left[g_1(\Sbf_1)+g_2(\Sbf_2)\right]$ which is also an obvious upper bound on the sum-rate capacity. Therefore, it is the sum-rate capacity.

If $\lambda_1>0$ and $\lambda_2>0$, we let
\bqa
\bar\gamma=\gamma,\quad \bar\lambda_i=\lambda_i,\quad \bar\eta_i=\eta_i\quad
\overline\Wbf_i=\Wbf_i-\lambda_j\Obf_i,\qquad i,j\in\{1,2\},i\neq j
\label{eq:lagrangianMultipliers}
\eqa
and since $\Sbf_i^*\Obf_i=\0bf$, $i=1,2$, then the KKT conditions (\ref{eq:KKTrateSu})-(\ref{eq:KKTWSu}) for the upper bound (\ref{eq:MIMOsumu}) are also satisfied. By the convexity of (\ref{eq:MIMOsumu}), $\Sbf_1^*$ and $\Sbf_2^*$ also maximize problem (\ref{eq:MIMOsumu}). Furthermore, problems (\ref{eq:MIMOsuml}) and (\ref{eq:MIMOsumu}) have the same maximum by the fact that $g_{si}\left(\Sbf_1^*,\Sbf_2^*\right)=\bar g_{si}\left(\Sbf_1^*,\Sbf_2^*\right)$, $i=1,2$. Therefore, the lower and upper bounds on the sum-rate capacity converge at $\left(\Sbf_1^*,\Sbf_2^*\right)$.

If $\lambda_1>0$ and $\lambda_2=0$, then we remove the constraint $R_1+R_2\leq \bar g_{s2}(\Sbf_1,\Sbf_2)$ in problem (\ref{eq:MIMOsumu}). Consequently, in Lemma \ref{lemma:MIMOSu}, we need the existence of only $\Abf_2$ to satisfy (\ref{eq:Markov}) and (\ref{eq:A}). The corresponding KKT conditions in (\ref{eq:KKTrateSu})-(\ref{eq:KKTWSu}) are changed into those equivalent to letting $\bar\lambda_2=0$. Then we can still choose the Lagrangian multipliers as in (\ref{eq:lagrangianMultipliers}). Therefore, $\Sbf_1^*$ and $\Sbf_2^*$ also maximize problem (\ref{eq:MIMOsumu}). Problems (\ref{eq:MIMOsuml}) and (\ref{eq:MIMOsumu}) have the same maximum which is also the sum-rate capacity.

The case for $\lambda_1=0$ and $\lambda_2>0$ is similarly proved by removing the constraint $R_1+R_2\leq \bar g_{s1}(\Sbf_1,\Sbf_2)$ from problem (\ref{eq:MIMOsumu}).
\epf
\begin{remark}
In the proof of Theorem \ref{theorem:MIMOs}, we remove the constraint $R_1+R_2\leq\bar g_{s2}(\Sbf_1,\Sbf_2)$ when $\lambda_2=0$ only because we do not need the existence of $\Abf_2$ to satisfy (\ref{eq:Markov}) and (\ref{eq:A}) which imply $g_{s2}\left(\Sbf_1^*,\Sbf_2^*\right)=\bar g_{s2}\left(\Sbf_1^*,\Sbf_2^*\right)$. Since the rate constraint $g_{s2}$ is inactive in the inner bound when $\lambda_2=0$ we can simply remove the constraint $\bar g_{s2}$ from the outer bound.
\end{remark}
\begin{theorem}
Suppose $\Sbf_i^*$, $i=1,2$ are maximizers of problem (\ref{eq:MIMOIB}) for a given $r\in\left[0,\max\frac{1}{2}\log\left|\Ibf+\Hbf_2\Sbf_2\Hbf_2^T\right|\right]$. For $i=1,2$, let $\beta_i$ and $\Kbf_i$ be the corresponding Lagrangian multipliers satisfying (\ref{eq:KKTrateI})-(\ref{eq:KKTK0}). For any $\beta_j>0$, $j=1,2$, if there exist $\Abf_i$, $i=1,2$, $i\neq j$, that satisfies (\ref{eq:Markov}) and (\ref{eq:A}) and
\bqa
\Kbf_i\succeq \beta_j\Obf_i
\eqa
where $\Obf_j$ is defined in (\ref{eq:O}), then the rate pair $\left\{R_1=R_1^*(r),R_2=r\right\}$ is on the boundary of the capacity region, and is achieved by the input distributions $\xp_i\sim\Nmat\left(\0bf,\Sbf_i^*\right)$, $i=1,2$, and jointly decoding the signal and the interference.
\label{theorem:MIMOB}
\end{theorem}
\bpf
The proof is similar to the proof of Theorem \ref{theorem:MIMOs}. We first modify problem (\ref{eq:MIMOoB}) according to $\beta_j$. If $\beta_j=0$, then we remove the constraint $R_1\leq\bar g_{sj}-r$.

By choosing
\bqn
\bar\alpha=\alpha,\quad \bar\beta_i=\beta_i,\quad \bar\nu_i=\eta_i\quad
\overline\Kbf_i=\Kbf_i-\beta_j\Obf_i,\qquad i,j\in\{1,2\},i\neq j
\eqn
then the KKT conditions in (\ref{eq:KKTrateO})-(\ref{eq:KKTOK0}) for the modified problem (\ref{eq:MIMOoB}) are satisfied. Therefore, the modified problem (\ref{eq:MIMOoB}) is also maximized at $\Sbf_1^*$ and $\Sbf_2^*$. Problems (\ref{eq:MIMOIB}) and the modified (\ref{eq:MIMOoB}) have the same maximum by the fact $g_{sj}\left(\Sbf_1^*,\Sbf_2^*\right)=\bar g_{sj}\left(\Sbf_1^*,\Sbf_2^*\right)$ for any $j$ with $\beta_j>0$.
\epf
\begin{remark}
Theorem \ref{theorem:MIMOB} is used to establish the boundary of the capacity region. For each boundary point, we need to find the corresponding matrices $\Abf_i$ satisfying (\ref{eq:Markov}) and (\ref{eq:A}) which gives one outer bound. This outer bound is tight at this particular point. Therefore, to find the whole capacity region, we need to find the tight outer bound for each boundary point. There are cases in which only part of the boundary points can be determined by Theorem \ref{theorem:MIMOB}, see Example \ref{example:partial}.
\end{remark}
\begin{remark}
In Theorems \ref{theorem:MIMOs} and \ref{theorem:MIMOB}, in case of $\lambda_j\neq 0$ or $\beta_j\neq 0$, we always need the existence of matrix $\Abf_i$, $i\neq j$, satisfying (\ref{eq:Markov}) and (\ref{eq:A}) even if $\Obf_i=\0bf$. The reason is that the corresponding tight outer bound can be established only when such $\Abf_i$ exists.
\end{remark}

\begin{remark}
If the conditions in Theorems \ref{theorem:MIMOs} and \ref{theorem:MIMOB} are satisfied, then the MIMO IC has generally strong interference at the sum-rate capacity or at the rate pair $\left\{R_1^*(r),r\right\}$. In both cases, the capacity is achieved by Gaussian input sequences and jointly decoding the signal and the interference. We show in the following that under conditions (\ref{eq:Markov}) and (\ref{eq:A}), inequalities (\ref{eq:strong1}) and (\ref{eq:strong2}) are satisfied for the input distribution $\xp_i^*\sim\Nmat\left(\0bf,\Sbf_i^*\right)$, $i=1,2$:
\bqa
I\left(\xp_1^*;\yp_1\left|\hspace{.05in}\xp_2^*\right.\right)
&{}={}&I\left(\xp_1^*;\Hbf_1\xp_1^*+\Fbf_2\xp_2^*+\zp_1\left|\hspace{.05in}\xp_2^*\right.\right)\nn\\
&{}={}&I\left(\xp_1^*;\Hbf_1\xp_1^*+\zp_1\right)\nn\\
&{}={}&\frac{1}{2}\log\left|\Ibf+\Hbf_1\Sbf_1^*\Hbf_1^T\right|\nn\\
&{}\stackrel{(a)}={}&\frac{1}{2}\log\left|\Ibf+\Hbf_1\Sbf_1^*\Fbf_1^T\Abf_1^T\right|\nn\\
&{}\stackrel{(b)}={}&\frac{1}{2}\log\left|\Ibf+\Abf_1\Fbf_1\Sbf_1^*\Fbf_1^T\Abf_1^T\right|\nn\\
&{}={}& I\left(\xp_1^*;\Abf_1\Fbf_1\xp_1^*+\zp_1\right)\nn\\
&{}\stackrel{(c)}={}& I\left(\xp_1^*;\Abf_1\left(\Fbf_1\xp_1^*+\zp_1\right)+\left(\Ibf-\Abf_1\Abf_1^T\right)\tilde\zp\right)\nn\\
&{}\leq{}& I\left(\xp_1^*;\Abf_1\left(\Fbf_1\xp_1^*+\zp_1\right)+\left(\Ibf-\Abf_1\Abf_1^T\right)\tilde\zp,\tilde\zp\right)\nn\\
&{}={}& I\left(\xp_1^*;\Abf_1\left(\Fbf_1\xp_1^*+\zp_1\right)\right)\nn\\
&{}\stackrel{(d)}\leq{}& I\left(\xp_1^*;\Fbf_1\xp_1^*+\zp_1\right)\nn\\
&{}={}& I\left(\xp_1^*;\yp_2\left|\hspace{.05in}\xp_2^*\right.\right)
\label{eq:satisfy}
\eqa
where (a) is by (\ref{eq:Markov}); (b) is also by (\ref{eq:Markov}) which implies $\Hbf_1\Sbf_1^*=\Abf_1\Fbf_1\Sbf_1^*$; (c) is by (\ref{eq:A}) and we let $\tilde\zp\sim\Nmat\left(\0bf,\Ibf\right)$ be independent of $\xp_1^*$ and $\zp_1$; and (d) is by the Markov relationship $\xp_1^*\rightarrow\xp_1^*+\zp_1\rightarrow\Abf_1\left(\xp_1^*+\zp_1\right)$. Similarly, we can show $I\left(\xp_2^*;\yp_2\left|\hspace{.05in}\xp_1^*\right.\right)\leq I\left(\xp_2^*;\yp_1\left|\hspace{.05in}\xp_1^*\right.\right)$. Therefore, the strong interference conditions (\ref{eq:strong1}) and (\ref{eq:strong2}) are both satisfied for a MIMO IC with generally strong interference at the capacity achieving input distributions. For other input distributions, the MIMO IC with generally strong interference may not satisfy the strong interference conditions (\ref{eq:strong1}) and (\ref{eq:strong2}).
\label{remark:satisfyS}
\end{remark}

\begin{remark}
If an MIMO IC has generally strong interference at rate pair $\{R_1,R_2\}$ and satisfies the conditions in Theorem \ref{theorem:MIMOB}, then this rate pair is in the achievable region given in (\ref{eq:MIMOhk}) by replacing $\Sbf_i$ with $\Sbf_i^*$, for $i=1,2$. By Remark \ref{remark:satisfyS}, we have
\bqn
R_1\leq\frac{1}{2}\log\left|\Ibf+\Fbf_1\Sbf_1^*\Fbf_1^T\right|\\
R_2\leq\frac{1}{2}\log\left|\Ibf+\Fbf_2\Sbf_2^*\Fbf_2^T\right|.
\eqn
On combining the above constraints with those in (\ref{eq:MIMOhk}), we have
\bqn
0\leq R_1&{}\leq {}& \min\left\{I\left(\xp_1^*,\yp_1\left|\hspace{.05in}\xp_2^*\right.\right),I\left(\xp_1^*,\yp_2\left|\hspace{.05in}\xp_2^*\right.\right)\right\}\\
0\leq R_2&{}\leq {}& \min\left\{I\left(\xp_2^*,\yp_2\left|\hspace{.05in}\xp_1^*\right.\right),I\left(\xp_2^*,\yp_2\left|\hspace{.05in}\xp_1^*\right.\right)\right\}\\
R_1+R_2&{}\leq {}&
\min\left\{I\left(\xp_1^*\xp_2^*,\yp_1\right),I\left(\xp_1^*\xp_2^*,\yp_2\right)\right\}.
\eqn
The above region is the same as the achievable region of a compound multiple access channel (by requiring both receivers to correctly decode messages from both transmitters). Therefore, under generally strong interference, the receivers can still correctly decode the interference for the capacity achieving distribution.
\end{remark}

\begin{remark}
Theorems \ref{theorem:MIMOs} and \ref{theorem:MIMOB} specify the sum-rate capacity and the boundary points of the capacity region for a MIMO IC with generally strong interference. The conditions of Theorems \ref{theorem:MIMOs} and \ref{theorem:MIMOB} require the optimization of problems (\ref{eq:MIMOsuml}) and (\ref{eq:MIMOIB}) and the solution of (\ref{eq:Markov}) for matrices $\Abf_1$ and $\Abf_2$. Since both (\ref{eq:MIMOsuml}) and (\ref{eq:MIMOIB}) are convex optimization problems, they can be efficiently solved using standard optimization algorithms. Equation (\ref{eq:Markov}) for matrices $\Abf_1$ and $\Abf_2$ is a special case of the Sylvester equation \cite{Bartels&Stewart:72ACM}. Once $\Sbf_1^*$ and $\Sbf_2^*$ are obtained, the matrices $\Abf_1$ and $\Abf_2$ can be obtained by solving the following linear equations\cite[Remark 7]{Shang&Poor:11IT_submission_MIMO}:
\bqa
&&\Ibf\otimes\left(\Sbf_1^*\Fbf_1^T\right)\Vec(\Abf_1)=
\Vec\left(\Sbf_1^*\Hbf_1^T\right)\nn\\
&&\Ibf\otimes\left(\Sbf_2^*\Fbf_2^T\right)\Vec(\Abf_2)=
\Vec\left(\Sbf_2^*\Hbf_2^T\right)\nn
\eqa
Therefore, the existence of $\Abf_1$ and $\Abf_2$ can be determined by the theory of linear equations. Once $\Sbf_1^*$, $\Sbf_2^*$, $\Abf_1$ and $\Abf_2$ are obtained, the Lagrangian multipliers $\lambda_i$, $\Wbf_i$, $\beta_i$ and $\Kbf_i$, $i=1,2$, can be obtained by solving the KKT conditions. Therefore, Theorems \ref{theorem:MIMOs} and \ref{theorem:MIMOB} can be efficiently applied to any MIMO IC.
\end{remark}
\begin{remark}
If the strong interference conditions (\ref{eq:mimos1}) and (\ref{eq:mimos2}) are satisfied, we have $\Obf_i=\0bf$, $i=1,2$. Therefore, the generally strong interference conditions are automatically satisfied. Furthermore, for the very strong interference we have $\beta_1=\beta_2=0$ when $r=\frac{1}{2}\max_{\Sbf_2}\log\left|\Ibf+\Hbf_2\Sbf_2\Hbf_2^T\right|$. Therefore, the generally strong interference conditions are also satisfied and we do not need the existence of $\Abf_1$ or $\Abf_2$.
\end{remark}

In the following, we apply Theorems \ref{theorem:MIMOs} and \ref{theorem:MIMOB} to SIMO and MISO ICs and derive their capacity region under generally strong interference.

\section{SIMO ICs}
The received signals of a SIMO IC can be written as
\bqa
\yp_1&{}={}&X_1\hp_1+X_2\fp_2+\zp_1\nn\\
\yp_2&{}={}&X_2\hp_2+X_1\fp_1+\zp_2.
\eqa
where $\hp_i$ and $\fp_i$ $i=1,2$, are both $t_i\times 1$ column vectors. We need to find $t_i\times t_i$ matrices $\Abf_i$ that satisfy (\ref{eq:Markov}) and (\ref{eq:A}). Since the $\Sbf_i^*$'s are now scalars, we have
\bqa
\Abf_i=\frac{\hp_i\rhobf_i^T}{\rhobf_i^T\fp_i},\quad i=1,2
\label{eq:simoA}
\eqa
where $\rhobf_i$ is a nonzero $t_i\times 1$ column vector. For condition (\ref{eq:A}), we need
\bqa
\Ibf\succeq\Abf_i\Abf_i^T=\frac{\hp_i\rhobf_i^T\rhobf_i\hp_i^T}{\left(\rhobf_i^T\fp_i\right)^2}
=\frac{\hp_i\hp_i^T}{\left\|\fp_i\right\|^2\cos^2\angle\left(\rhobf_i,\fp_i\right)}\quad i=1,2.
\eqa
By \cite[Lemma 6 by $\Bbf=\Ibf$]{Shang-etal:10IT_mimo} the above condition is equivalent to
\bqa
\|\hp_i\|^2\leq\|\fp_i\|^2\cos^2\angle\left(\rhobf_i,\fp_i\right)\leq\|\fp_i\|^2,\quad i=1,2.
\eqa
On the other hand, we have $\Obf_i=\0bf$, $i=1,2$, by (\ref{eq:simoA}). Therefore, the SIMO IC has generally strong interference for the entire capacity region if for $i=1,2$, $\|\hp_i\|\leq\|\fp_i\|$ for any $\fp_i\neq\0bf$. This condition is the same as that in \cite{Vishwanath&Jafar:04ITW} and is also included as a special case of \cite{Shang-etal:10IT_mimo}, i.e., the generally strong interference obtained from Theorems \ref{theorem:MIMOs} and \ref{theorem:MIMOB} is exactly the same as strong interference.

It is straightforward to show that the very strong interference condition (\ref{eq:Gvstrong})
\bqa
\log\left|\Ibf+P_i\hp_i\hp_i^T\right|\leq\log\left|\Ibf+P_i\fp_i\fp_i^T+P_j\hp_j\hp_j^T\right|-\log\left|\Ibf+P_j\hp_j\hp_j^T\right|
\quad i,j\in\{1,2\}, i\neq j
\eqa
is equivalent to
\bqa
\frac{\|\fp_i\|^2}{\|\hp_i\|^2}\geq\frac{1+P_j\|\hp_j\|^2}{1+P_j\|\hp_j\|^2\sin^2\angle(\fp_i,\hp_j)},\quad i,j\in\{1,2\},i\neq j.
\eqa
Therefore, for the SIMO IC the very strong interference condition is a special case of the (generally) strong interference condition.

\section{MISO ICs}
In this section, we use the MISO IC as an example to show how Theorems \ref{theorem:MIMOs} and \ref{theorem:MIMOB} are applied to obtain its capacity region under the generally strong interference. The received signals of a MISO IC are defined as
\bqa
\hat Y_1&{}={}&\hat\hp_1^T\hat\xp_1+\hat\fp_2^T\hat\xp_2+Z_1\nn\\
\hat Y_2&{}={}&\hat\hp_2^T\hat\xp_2+\hat\fp_1^T\hat\xp_1+Z_2
\label{eq:miso_ori}
\eqa
where $\hat\hp_i$ and $\hat\fp_i$, $i=1,2$, are $t_i\times 1$ channel vectors, $Z_i\sim\Nmat(0,1)$ and
\bqa
\sum_{j=1}^n\tr\left(E\left[\hat\xp_{ij}\hat\xp_{ij}^T\right]\right)\leq n\hat P_i,\quad i=1,2.
\eqa
It has been shown that the capacity region of channel (\ref{eq:miso_ori}) is the same as that of a MISO IC with only two transmit antennas \cite{Annapureddy&Veeravalli:11IT}. In fact, the capacity region of an $m$-user MISO IC is the same as that of an $m$-user MISO IC with each $i$th transmitter having $\min\{t_i,m\}$ antennas. The reduction process of transmitter antennas is shown in  \cite[eqs.(45)-(47)]{Shang-etal:11IT_miso} and its application to the two-user MISO IC is shown in \cite[eqs.(78)-(83)]{Shang&Poor:11IT_submission_MIMO}. We rewrite the result of \cite{Shang&Poor:11IT_submission_MIMO} as follows: channel (\ref{eq:miso_ori}) is equivalent to the MISO IC defined as
\bqa
Y_1&{}={}&\hp_1^T\xp_1+\fp_2^T\xp_2+Z_1\nn\\
Y_2&{}={}&\hp_2^T\xp_2+\fp_1^T\xp_1+Z_2
\label{eq:miso}
\eqa
where, for $i=1,2$,
\bqa
\hp_i&{}={}&\left[\begin{array}{c}
              \cos\theta_i \\
              \sin\theta_i
            \end{array}
\right]\\
\fp_i&{}={}&\left[\begin{array}{c}
              \sqrt{a_i} \\
              0
            \end{array}
\right]
\eqa
and
\bqa
\theta_i&{}={}&\angle\left(\hat\hp_i,\hat\fp_i\right)\\
a_i&{}={}&\frac{\left\|\hat\fp_i\right\|^2}{\left\|\hat\hp_i\right\|^2}.
\eqa
The power constraint is now
\bqa
\sum_{j=1}^n\tr\left(E\left[\xp_{ij}\xp_{ij}^T\right]\right)\leq n P_i=n\hat P_i\left\|\hat\hp_i\right\|^2,\quad i=1,2.
\eqa
If $\Sbf_i$ is the input covariance matrix of user $i$ for equivalent channel (\ref{eq:miso}), the corresponding input covariance matrix $\hat\Sbf_i$ for the original channel is obtained in \cite[eq. (88)]{Shang&Poor:11IT_submission_MIMO}. In the sequel, we use (\ref{eq:miso}) as the channel model for MISO ICs.

We first obtain the joint decoding achievable rate region given in Lemma \ref{lemma:MIMOhk}.
\begin{lemma}
The achievable rate region (\ref{eq:MIMOhk}) for a MISO IC is
\bqa
\bigcup_{\phi_i\in\left[0,\frac{\pi}{2}\right]}\left\{\begin{array}{c}
    R_1\leq \frac{1}{2}\log\left(1+P_1\sin^2(\theta_1+\tau_1\phi_1)\right) \\
    R_2\leq \frac{1}{2}\log\left(1+P_2\sin^2(\theta_2+\tau_2\phi_2)\right) \\
    R_1+R_2\leq \frac{1}{2}\log\left(1+P_1\sin^2(\theta_1+\tau_1\phi_1)+a_2P_2\sin^2\phi_2\right) \\
    R_1+R_2\leq \frac{1}{2}\log\left(1+P_2\sin^2(\theta_2+\tau_2\phi_2)+a_1P_1\sin^2\phi_1\right)
                                                      \end{array}
\right\}
\label{eq:misoI}
\eqa
where $\tau_i=\textrm{sign}(\cos(\theta_i))$, and is achieved by
\bqa
\Sbf_i=P_i\left[\begin{array}{c}
              \sin\phi_i \\
              \tau_i\cos\phi_i
            \end{array}
\right]\left[\begin{array}{c}
              \sin\phi_i \\
              \tau_i\cos\phi_i
            \end{array}
\right]^T=P_i\left[\begin{array}{cc}
                     \sin^2\phi_i &\quad \tau_i\cos\phi_i\sin\phi_i \\
                     \tau_i\cos\phi_i\sin\phi_i &\quad \cos^2\phi_i
                   \end{array}
\right],\quad i=1,2.
\label{eq:bfs}
\eqa
\label{lemma:misoI}
\end{lemma}
\bpf
It has been shown in \cite[Lemma 2]{Shang-etal:11IT_miso} that given
\bqa
\fp_i^T\Sbf_i\fp_i=a_iP_i\sin^2\phi_i,\qquad \phi\in\left[0,\frac{\pi}{2}\right]
\label{eq:bfcond}
\eqa
we have
\bqa
\hp_i^T\Sbf_i\hp_i\leq P_i\sin^2(\theta_i+\tau_i\phi_i)
\label{eq:bfmax}
\eqa
and the equality is achieved by (\ref{eq:bfs}). Therefore, region (\ref{eq:MIMOhk}) reduces to (\ref{eq:misoI}).
\epf

Lemma \ref{lemma:misoI} reveals the fact that all the boundary points of the rate region (\ref{eq:MIMOhk}) can be achieved by rank-1 beamforming. Therefore, to determine whether the boundary points of region (\ref{eq:MIMOhk}) are also the boundary points of the capacity region, we need to consider only the rank-1 covariance matrices. By Theorems \ref{theorem:MIMOs} and \ref{theorem:MIMOB}, we obtain the sum-rate capacity and the boundary of the capacity region in the following propositions.
\begin{proposition}
For a MISO IC defined in (\ref{eq:miso_ori}) and its equivalent channel (\ref{eq:miso}), let $\Sbf_i^*$, $i=1,2$, be optimal for problem (\ref{eq:MIMOsuml}) where $\Hbf_i=\hp_i^T$ and $\Fbf_i=\fp_i^T$, $i=1,2$; then there exist $\phi_i^*\in\left[0,\frac{\pi}{2}\right]$, $i=1,2$, such that
\bqa
\Sbf_i^*=P_i\left[\begin{array}{cc}
                     \sin^2\phi_i^* &\quad \tau_i\cos\phi_i^*\sin\phi_i^* \\
                     \tau_i\cos\phi_i^*\sin\phi_i^* &\quad \cos^2\phi_i^*
                   \end{array}
\right]
\label{eq:misoSstar}
\eqa
where $\tau_i=\textrm{sign}(\cos\theta_i)$. Furthermore, let $\lambda_i$ and $\Wbf_i$, $i=1,2$, be the Lagrangian multipliers satisfying (\ref{eq:KKTrateSl})-(\ref{eq:KKTW0}). For any $\lambda_j>0$, $j=1,2,j\neq i$, if
\bqa
&&\sin^2\left(\theta_i+\tau_i\phi_i^*\right)< a_i\sin^2\phi_i^*
\label{eq:misoScond1}\\
&&\Wbf_i\succeq \frac{\lambda_j}{2}\cdot\frac{a\sin^2\theta_i}{a\sin^2\phi_i^*-\sin^2\left(\theta_i+\phi_i^*\right)}
\left[\begin{array}{cc}
        \cos^2\phi_i^* & \quad -\tau_i\sin\phi_i^*\cos\phi_i^* \\
        -\tau_i\sin\phi_i^*\cos\phi_i^* &\quad \sin^2\phi_i^*
      \end{array}
\right]
\label{eq:misoScond2}
\eqa
then the sum-rate capacity is the maximum in (\ref{eq:MIMOsuml}) and is achieved by Gaussian inputs $\xp_i\sim\Nmat\left(\0bf,\Sbf_i^*\right)$ and by jointly decoding the signal and the interference.
\label{prop:misoSC}
\end{proposition}
\bpf
The fact that the optimal $\Sbf_i^*$'s have the form in (\ref{eq:misoSstar}) is determined by (\ref{eq:bfcond}) and (\ref{eq:bfmax}). By Theorem \ref{theorem:MIMOs}, the maximum in (\ref{eq:MIMOsuml}) is the sum-rate capacity, if for any $\lambda_j>0$, $j=1,2,j\neq i$ the following conditions are satisfied:
\bqa
&&\left[\begin{array}{cc}
        1 & \quad A_i \\
        A_i & \quad 1
      \end{array}
\right]\succeq\0bf
\label{eq:misoGenie}\\
&&\Sbf_i^*\hp_i=\Sbf_i^*\fp_iA_i
\label{eq:misoMarkov}\\
&&\Wbf_i\succeq\lambda_j\Obf_i=\frac{\lambda_j}{2\left(1-A_i^2\right)}\left(\hp_i-A_i\fp_i\right)\left(\hp_i-A_i\fp_i\right)^T,\quad i,j\in\{1,2\},i\neq j.
\label{eq:misoOpt}
\eqa
Since $\Sbf_i^*$ is a unit-rank matrix, there always exists a scalar $A_i$ that satisfies (\ref{eq:misoMarkov}), and
\bqa
A_i=\frac{\tau_i\sin\left(\theta_i+\tau_i\phi_i^*\right)}{\sqrt{a_i}\sin\phi_i^*}.
\label{eq:misoAi}
\eqa
With (\ref{eq:misoAi}), conditions (\ref{eq:misoGenie})-(\ref{eq:misoOpt}) reduce to (\ref{eq:misoScond1}) and (\ref{eq:misoScond2}).
\epf
\begin{proposition}
For a MISO IC defined in (\ref{eq:miso_ori}) and its equivalent channel (\ref{eq:miso}), let $\Sbf_i^*$, $i=1,2$, be optimal for problem (\ref{eq:MIMOIB}) for a given $r\in\left[0,\frac{1}{2}\log(1+P_2)\right]$ where $\Hbf_i=\hp_i^T$ and $\Fbf_i=\fp_i^T$, $i=1,2$; then for $i=1,2$, there exist $\phi_i^*\in\left[0,\frac{\pi}{2}\right]$ such that
\bqa
\Sbf_i^*=P_i\left[\begin{array}{cc}
                     \sin^2\phi_i^* &\quad \tau_i\cos\phi_i^*\sin\phi_i^* \\
                     \tau_i\cos\phi_i^*\sin\phi_i^* &\quad \cos^2\phi_i^*
                   \end{array}
\right]
\label{eq:misoSstarB}
\eqa
where $\tau_i=\textrm{sign}(\cos\theta_i)$. Furthermore, let $R_1^*(r)$ be the maximum in problem (\ref{eq:MIMOIB}), and let $\beta_i$ and $\Kbf_i$, $i=1,2$, be the Lagrangian multipliers satisfying (\ref{eq:KKTrateI})-(\ref{eq:KKTK0}). For any $\beta_j>0$, $j=1,2,j\neq i$, if
\bqa
&&\sin^2\left(\theta_i+\tau_i\phi_i^*\right)< a_i\sin^2\phi_i^*
\label{eq:misoSBcond1}\\
&&\Kbf_i\succeq \frac{\beta_j}{2}\cdot\frac{a\sin^2\theta_i}{a\sin^2\phi_i^*-\sin^2\left(\theta_i+\phi_i^*\right)}
\left[\begin{array}{cc}
        \cos^2\phi_i^* & \quad -\tau_i\sin\phi_i^*\cos\phi_i^* \\
        -\tau_i\sin\phi_i^*\cos\phi_i^* &\quad \sin^2\phi_i^*
      \end{array}
\right]
\label{eq:misoSBcond2}
\eqa
then the rate pair $\left(R_1^*\left(r\right),r\right)$ is on the boundary of the capacity region, and is achieved by Gaussian inputs $\xp_i\sim\Nmat\left(\0bf,\Sbf_i^*\right)$ and by fully decoding the interference.
\label{prop:misoCB}
\end{proposition}
\bpf
The proof is identical to that of Proposition \ref{prop:misoSC} and hence is omitted.
\epf

Propositions \ref{prop:misoSC} and \ref{prop:misoCB} provide sufficient conditions for a MISO IC to have generally strong interference. Those conditions are more amenable to numerical evaluation since the optimal input covariance matrices $\Sbf_i^*$ can be obtained using standard convex optimization algorithms, while analytical closed-form expressions for $\Sbf_i^*$ are difficult to derive in general except in the very strong interference case:
\begin{proposition}
For the MISO IC if $a_i=0$ or $a_i\cos^2\theta_i\geq 1+P_i$, $i=1,2$, then the capacity region is $0\leq R_i\leq \frac{1}{2}\log(1+P_i)$, $i=1,2$, and is achieved by choosing $\xp_i\sim\Nmat\left(\0bf,\Sbf_i^*\right)$, $i=1,2$, where
\bqa
\Sbf_i^*=P_i\left[\begin{array}{cc}
                    \cos^2\theta_i & \quad \tau_i\cos\theta_i\sin\theta_i \\
                    \tau_i\cos\theta_i\sin\theta_i & \quad \sin^2\theta_i
                  \end{array}
\right]
\eqa
and $\tau_i=\sign\left(\cos\theta_i\right)$.
\end{proposition}

The proof is straightforward and hence is omitted.

 In the following, we apply these two propositions to two special cases of MISO ICs: the MISO ZIC with $\fp_1=0$, and the symmetric MISO IC with $\theta_1=\theta_2\neq\frac{\pi}{2}$, $a_1=a_2> 0$ and $P_1=P_2>0$.

\subsection{MISO ZIC}

A MISO ZIC is defined as in (\ref{eq:miso_ori}) with $\hat\fp_1=\0bf$. By using (\ref{eq:miso}), the capacity region of such a MISO IC is equivalent to the channel defined as
\bqa
Y_1&{}={}&X_1+\fp^T\xp_2+Z_1\nn\\
Y_2&{}={}&\hp^T\xp_2+Z_2
\label{eq:eqZ}
\eqa
where we let $\theta_1=\angle(\hp_1,\fp_1)=0$ when $\fp_1=\0bf$. Therefore, $\xp_1$ reduces to a scalar $X_1$. The power constraints are still $P_1$ and $P_2$ for users $1$ and $2$, respectively.

When $a=0$ or $\theta=\frac{\pi}{2}$, the capacity region of this MISO ZIC is trivially obtained. When $\theta\in\{0,\pi\}$, the MISO ZIC reduces to a scalar Gaussian ZIC of which the capacity region under (generally) strong interference has been obtained. Without loss of generality, we assume $a\neq 0$ and $\theta\notin\left\{ 0,\frac{\pi}{2},\pi\right\}$ in the sequel.

We obtain the joint decoding achievable region of this MISO ZIC by Lemma \ref{lemma:misoI}.

\begin{lemma}
For a MISO ZIC defined in (\ref{eq:eqZ}), the achievable rate region (\ref{eq:misoI}) is
\bqa
\bigcup_{\phi\in\left[0,\frac{\pi}{2}\right]}\left\{\begin{array}{l}
    R_1\leq\frac{1}{2}\log(1+P_1) \\
    R_2\leq\frac{1}{2}\log\left(1+P_2\sin^2\left(\theta+\tau\phi\right)\right) \\
    R_1+R_2\leq\frac{1}{2}\log\left(1+P_1+aP_2\sin^2\phi\right)
                                                    \end{array}
\right\}
\label{eq:MISOZhk}
\eqa
where $\tau=\textrm{sign}\left(\cos\theta\right)$.
\label{lemma:MISOZi}
\end{lemma}
\bpf
For a MISO ZIC with $\fp_1=\0bf$, the second receiver has no interference. Therefore, the second constraint on $R_1+R_2$ in (\ref{eq:misoI}) is not necessary and is hence removed.
\epf

Using Lemma \ref{lemma:MISOZi}, we obtain the largest sum rate and the boundary of the region defined in (\ref{eq:MISOZhk}) respectively in the following two lemmas.

\begin{lemma}
The largest sum rate of the region defined in (\ref{eq:MISOZhk}) is
\bqa
&&R_1+R_2\nn\\
&&=\left\{\begin{array}{ll}
                 \frac{1}{2}\log(1+P_1)+\frac{1}{2}\log(1+P_2) &\quad \textrm{if }\cos^2\theta\geq\frac{1+P_1}{a}  \\
                 \frac{1}{2}\log(1+P_1+aP_2) &\quad \textrm{if }\cos^2\theta\geq\frac{a}{1+P_1}  \\
                 \frac{1}{2}\log\left(1+P_1+aP_2\sin^2\phi_{ez}\right)
                 =\frac{1}{2}\log(1+P_1)\\
                 \hspace{1.8in}+\frac{1}{2}\log\left(1+P_2\sin^2\left(\theta+\tau\phi_{ez}\right)\right)&\quad \textrm{if }\cos^2\theta
                 \leq\min\left\{\frac{a}{1+P_1},\frac{1+P_1}{a}\right\}
               \end{array}
\right.
\eqa
where $\tau=\sign\left(\cos\theta\right)$ and
\bqa
\phi_{ez}=\textrm{atan}\frac{\sin\theta}{\sqrt\frac{a}{1+P_1}-\tau\cdot\cos\theta}.
\label{eq:phiStar}
\eqa
The corresponding $\Sbf$ that achieves the sum rate is
\begin{subequations}
\begin{numcases}{\Sbf^*=}
                 P_2\left[\begin{array}{cc}
                         \cos^2\theta &\quad \tau\sin\theta\cos\theta \\
                         \tau\sin\theta\cos\theta &\quad \sin^2\theta
                       \end{array}
                 \right] \hspace{.83in} \textrm{if }\cos^2\theta\geq\frac{1+P_1}{a}
\label{eq:SmisoZsvs}\\
                 \left[\begin{array}{cc}
                         P_2 &\quad 0 \\
                         0 &\quad 0
                       \end{array}
                 \right] \hspace{2.23in} \textrm{if }\cos^2\theta\geq\frac{a}{1+P_1}
\label{eq:SmisoZss1}\\
                  P_2\left[\begin{array}{cc}
                         \sin^2\phi_{ez} &\quad \tau\sin\phi_{ez}\cos\phi_{ez} \\
                         \tau\sin\phi_{ez}\cos\phi_{ez} &\qquad \cos^2\phi_{ez}
                       \end{array}
                 \right]\qquad \textrm{if }\cos^2\theta
                 \leq\min\left\{\frac{a}{1+P_1},\frac{1+P_1}{a}\right\}.
\label{eq:SmisoZss2}
\end{numcases}
\end{subequations}
\label{lemma:MISOZsl}
\end{lemma}
\bpf
We consider the case of $\cos\theta\geq 0$, and consequently $\tau=1$. The case for $\cos\theta<0$ can be similarly proved. The sum rate for the achievable region given in Lemma \ref{lemma:MISOZi} is bounded as
\bqa
R_1+R_2&{}\leq{}& \frac{1}{2}\log(1+P_1)+\frac{1}{2}\max_{\phi\in\left[0,\frac{\pi}{2}\right]}\min\left\{
\log\left(1+P_2\sin^2(\theta+\phi)\right),\log\left(1+\frac{aP_2\sin^2\phi}{1+P_1}\right)\right\}\nn\\
&{}={}&\frac{1}{2}\log(1+P_1)+\frac{1}{2}\log\left(1+P_2\cdot\max_{\phi\in\left[0,\frac{\pi}{2}\right]}\min\left\{d_1(\phi),
d_2(\phi)\right\}\right)
\label{eq:maxD1D2}
\eqa
where
\bqa
d_1(\phi)&{}\triangleq{}&\sin^2(\theta+\phi)\\
d_2(\phi)&{}={}&\frac{a\sin^2\phi}{1+P_1}.
\eqa
When $\cos^2\theta\geq \frac{a}{1+P_1}$, we have $d_1\left(\phi\right)\geq d_2\left(\phi\right)$ for all $\phi$; therefore, $\phi=\frac{\pi}{2}$ maximizes (\ref{eq:maxD1D2}). When $\cos^2\theta<\frac{a}{1+P_1}$, we have
\bqa
\max_{\phi\in\left[0,\frac{\pi}{2}\right]}\left\{d_1(\phi),d_2(\phi)\right\}=\left\{\begin{array}{cc}
    d_1(\phi) &\quad \textrm{if }0\leq\phi\leq\phi_{ez} \\
    d_2(\phi) &\quad \textrm{if }\phi_{ez}\leq\phi\leq\frac{\pi}{2}
                                                                                    \end{array}
\right.
\eqa
where $\phi_{ez}$ is defined in (\ref{eq:phiStar}), which means that
\bqa
\sin^2\left(\theta+\phi_{ez}\right)=\frac{a\sin^2\phi_{ez}}{1+P_1}.
\label{eq:eqPhiStar}
\eqa
It can be shown that when $\cos^2\theta\geq\frac{1+P_1}{a}$, (\ref{eq:maxD1D2}) is maximized by $\phi=\frac{\pi}{2}-\theta$; and when $\cos^2\theta\leq\min\left\{\frac{1+P_1}{a},\frac{a}{1+P_1}\right\}$, (\ref{eq:maxD1D2}) is maximized by $\phi=\phi_{ez}$.
\epf

We then obtain the boundary of the region defined in Lemma \ref{lemma:MISOZi}.

\begin{lemma}
The following rate pairs are on the boundary of the region defined in (\ref{eq:MISOZhk}):
\begin{subequations}
\begin{numcases}{}
\left\{R_1=\frac{1}{2}\log(1+P_1),\quad R_2=\frac{1}{2}\log(1+P_2)\right\}\hspace{.725in}\textrm{if }\cos^2\theta\geq\frac{1+P_1}{a}
\label{eq:MISOzicVSi}\\
\bigcup_{\phi\in\left[\tau\left(\frac{\pi}{2}-\theta\right),\frac{\pi}{2}\right]}\left\{\begin{array}{l}
    R_1=\frac{1}{2}\log\left(1+P_1+aP_2\sin^2\phi\right)-R_2 \\
    R_2=\frac{1}{2}\log \left(1+P_2\sin^2\left(\theta+\tau\phi\right)\right)
                                                                                     \end{array}
\right\}\quad \textrm{if }\cos^2\theta\geq\frac{a}{1+P_1}
\label{eq:MISOzicSi1}\\
\bigcup_{\phi\in\left[\tau\left(\frac{\pi}{2}-\theta\right),\phi^*\right]}\left\{\begin{array}{l}
    R_1=\frac{1}{2}\log\left(1+P_1+aP_2\sin^2\phi\right)-R_2 \\
    R_2=\frac{1}{2}\log \left(1+P_2\sin^2\left(\theta+\tau\phi\right)\right)
                                                                                     \end{array}
\right\}\quad \textrm{if }\cos^2\theta
                 \leq\min\left\{\frac{a}{1+P_1},\frac{1+P_1}{a}\right\}\nn\\
\label{eq:MISOzicSi2}
\end{numcases}
\end{subequations}
where $\tau=\sign\left(\cos\theta\right)$ and $\phi_{ez}$ is defined in (\ref{eq:phiStar}). The corresponding $\Sbf$ that achieves these boundary points is

\begin{subequations}
\begin{numcases}{\Sbf^*=}
                 P_2\left[\begin{array}{cc}
                         \cos^2\theta &\quad \tau\sin\theta\cos\theta \\
                         \tau\sin\theta\cos\theta &\quad \sin^2\theta
                       \end{array}
                 \right] \hspace{.7in} \textrm{if }\cos^2\theta\geq\frac{1+P_1}{a}
\label{eq:misoZvsS}\\
                  P_2\left[\begin{array}{cc}
                         \sin^2\phi &\qquad \tau\sin\phi\cos\phi \\
                         \tau\sin\phi\cos\phi &\qquad \cos^2\phi
                       \end{array}
                 \right]\hspace{.5in} \textrm{otherwise }.
\label{eq:misoZsS}
\end{numcases}
\end{subequations}
\label{lemma:MISOZIB}
\end{lemma}
\bpf
It is obvious that when $\cos^2\theta\geq\frac{1+P_1}{a}$, the $R_1+R_2$ constraint becomes redundant by choosing $\tau\phi=\frac{\pi}{2}-\theta$ which maximizes $R_2$. Therefore, (\ref{eq:MISOzicVSi}) determines the boundary points. For the case of $\cos^2\theta\leq\frac{1+P_1}{a}$, we prove (\ref{eq:MISOzicSi1}) and (\ref{eq:MISOzicSi2}) for $\cos\theta\geq 0$. The results for $\cos\theta< 0$ can be proved similarly.

By Lemma \ref{lemma:MIMOIB}, for $R_2=r$, the maximal $R_1$ is determined by
\bqa
\max &&\quad R_1\nn\\
\textrm{subject to}&&\quad R_1\leq \frac{1}{2}\log(1+P_1)\nn\\
&&\quad R_2=r\nn\\
&&\quad R_2\leq \frac{1}{2}\log\left(1+P_2\sin^2\left(\theta+\omega\right)\right)\nn\\
&&\quad R_1\leq \frac{1}{2}\log\left(1+P_1+aP_2\sin^2\omega\right)-r\nn\\
&&\quad\omega\in\left[0,\frac{\pi}{2}\right].
\label{eq:MISOZiB1}
\eqa

By Lemma \ref{lemma:MISOZsl}, when $\cos^2\theta\geq\frac{a}{1+P_1}$, the sum rate (\ref{eq:MISOzicSi1}) can be achieved by choosing $\Sbf$ as (\ref{eq:SmisoZss1}). For this input covariance matrix $\Sbf$, the line segment connecting the following two points are on the boundary:
\bqa
&&\left(R_1=\frac{1}{2}\log(1+P_1), R_2=\frac{1}{2}\log\left(1+\frac{aP_2}{1+P_1}\right)\right)
\label{eq:point1}\\
&&\left(R_1=\frac{1}{2}\log(1+P_1+aP_2)-\frac{1}{2}\log\left(1+P_2\cos^2\theta\right), R_2=\frac{1}{2}\log\left(1+P_2\cos^2\theta\right)\right).
\label{eq:point2}
\eqa
Therefore, we need to consider only the boundary points with $\frac{1}{2}\log\left(1+P_2\cos^2\theta\right)\leq R_2\leq\frac{1}{2}\log\left(1+P_2\right)$. Let
\bqa
r=\frac{1}{2}\log\left(1+P_2\sin^2\left(\theta+\phi\right)\right),\quad \phi\in\left[\frac{\pi}{2}-\theta,\frac{\pi}{2}\right];
\eqa
then problem (\ref{eq:MISOZiB1}) becomes
\bqa
\max&&\quad \frac{1}{2}\log\left(1+P_1+aP_2\sin^2\omega\right)-r\nn\\
\textrm{subject to}&&\quad \sin^2\left(\theta+\omega\right)\geq \sin^2\left(\theta+\phi\right)\nn\\
&&\quad \omega\in\left[0,\frac{\pi}{2}\right].
\label{eq:MISOZsl2}
\eqa
We note that in this case the bound $R_1\leq \frac{1}{2}\log(1+P_1)$ is redundant because $\cos^2\theta\geq\frac{a}{1+P_1}$. It can be shown that the maximum in problem (\ref{eq:MISOZsl2}) is achieved when $\omega=\phi$. Therefore, the points given in (\ref{eq:MISOzicSi1}) are on the boundary.

When $\cos^2\theta\leq\min\left\{\frac{a}{1+P_1},\frac{1+P_1}{a}\right\}$, the sum-rate line segment defined in (\ref{eq:point1}) and (\ref{eq:point2}) shrinks to one point:
\bqa
\left(R_1=\frac{1}{2}\log(1+P_1),R_2=\frac{1}{2}\log\left(1+P_2\sin^2\left(\theta+\phi_{ez}\right)\right)
=\frac{1}{2}\log\left(1+\frac{aP_2\sin^2\phi_{ez}}{1+P_1}\right)\right).
\eqa
Therefore, we need to consider only the boundary points with $\frac{1}{2}\log\left(1+P_2\sin^2\left(\theta+\phi^*\right)\right)\leq R_2\leq\frac{1}{2}\log\left(1+P_2\right)$. Let
\bqa
r=\frac{1}{2}\log\left(1+P_2\sin^2\left(\theta+\phi\right)\right),\quad\textrm{and } \phi\in\left[\frac{\pi}{2}-\theta,\phi_{ez}\right];
\eqa
then problem (\ref{eq:MISOZiB1}) becomes (\ref{eq:MISOZsl2}), which is maximized also by $\omega=\phi$. Therefore, (\ref{eq:MISOzicSi2}) is the boundary.
\epf
\begin{lemma}
The capacity region of a MISO ZIC is outer bounded by
\bqa
\bigcup_{\tr(\Sbf)\leq P_2,\Sbf\succeq\0bf}\left\{
\begin{array}{l}
  R_1\leq\dfrac{1}{2}\log(1+P_1) \\
  R_2\leq\dfrac{1}{2}\log(1+\hp^T\Sbf\hp) \\
  R_1+R_2\leq\dfrac{1}{2}\log\left(1+\dfrac{P_1}{1+\fp^T\Sbf\fp}\right) +\dfrac{1}{2}\log\left|\Ibf+\Sbf\left(\fp\fp^T+
            \dfrac{(\hp-A\fp)(\hp-A\fp)^T}{1-A^2}\right)\right|
\end{array}
\right\}\nn\\
\label{eq:MISOZo}
\eqa
where $A$ can be any value satisfying $A^2< 1$.
\label{lemma:MISOZo}
\end{lemma}

\bpf
We choose
\bqa
\Ebf_1&{}={}&\Ibf\nn\\
\Ebf_2&{}={}&\left[\begin{array}{cc}
                     1 &\quad A \\
                     A &\quad 1
                   \end{array}
\right];
\eqa
then by (\ref{eq:O}), we have
\bqa
\Obf_2=\dfrac{(\hp-A\fp)(\hp-A\fp)^T}{2\left(1-A^2\right)}.
\label{eq:MISOZO2}
\eqa
By Lemma \ref{lemma:MIMOo} and substituting (\ref{eq:MISOZO2}) into (\ref{eq:MIMOo}), we have that (\ref{eq:MISOZo}) is an outer bound for the capacity region.
\epf

Next, we obtain the sum-rate capacity and the boundary of the capacity for a MISO ZIC with generally strong interference.

\begin{proposition}
For the MISO ZIC defined in (\ref{eq:eqZ}), if
\bqa
a\cos^2\theta\geq 1+P_1
\eqa
then the sum-rate capacity is
\bqa
R_1+R_2=\frac{1}{2}\log(1+P_1)+\frac{1}{2}\log(1+P_2)
\label{eq:misozsCond0}
\eqa
and is achieved by (\ref{eq:SmisoZsvs}). If
\bqa
%
&&0<\frac{1+P_1\sin^2\theta}{1-P_2\sin^2\theta}\leq a\leq (1+P_1)\cos^2\theta
\label{eq:MISOZScond1_2}
%
\eqa
then the sum-rate capacity is
\bqa
R_1+R_2=\frac{1}{2}\log\left(1+P_1+aP_2\right)
\label{eq:MISOZS1}
\eqa
and is achieved by (\ref{eq:SmisoZss1}). If
\bqa
&&\hspace{-.3in}\cos^2\theta\leq\min\left\{\frac{a}{1+P_1},\frac{1+P_1}{a}\right\}
\label{eq:MISOZScond2_1}\\
%
&&\hspace{-.3in}P_1\sqrt{\frac{a}{1+P_1}}\cdot\tau\cos\theta\geq\left(1-\sqrt{\frac{a}{1+P_1}}\cdot\tau\cos\theta\right)
\left(1+P_1+\frac{aP_2\sin^2\theta}{\dfrac{a}{1+P_1}+1-2\sqrt{\dfrac{a}{1+P_1}}\cdot\tau\cos\theta}\right)
\label{eq:MISOZScond2_2}
\eqa
then the sum-rate capacity is
\bqa
R_1+R_2&{}={}&\frac{1}{2}\log\left(1+P_1+aP_2\sin^2\phi_{ez}\right)\nn\\
&{}={}&\frac{1}{2}\log\left(1+P_1\right)+\frac{1}{2}\log\left(1+P_2\sin^2\left(\theta+\tau\phi_{ez}\right)\right)
\label{eq:MISOZS2}
\eqa
and is achieved by (\ref{eq:SmisoZss2}).
\label{prop:misozS}
\end{proposition}
\bpf
%
%
%
%
%
We consider only the case in which $\cos\theta\geq 0$, and consequently, $\tau=1$. The case for $\tau=-1$ can be similarly proved. When $a\cos^2\theta\geq 1+P_1$, the MISO IC has very strong interference. Its sum-rate capacity is trivially proved. Next, we first consider the case of  (\ref{eq:MISOZScond1_2}). Using Lemma \ref{lemma:MISOZsl}, the maximal sum rate of (\ref{eq:MISOZhk}) is (\ref{eq:MISOZS1}) and is achieved by (\ref{eq:SmisoZss1}). Then from Lemma \ref{lemma:MIMOSl} there exist Lagrangian multipliers that satisfy
\bsq
&&\gamma+\lambda_{1}=1
\label{eq:KKTrateSl_1}\\
&&\Wbf_2=-\frac{\gamma\hp\hp^T}{2\left(1+P_2\cos^2\theta\right)}-\frac{\lambda_{1}\fp\fp^T}{2\left(1+P_1+aP_2\right)}+\eta_2\Ibf
\label{eq:KKTW2_1}\\
&&\gamma=0
\label{eq:gamma_1}\\
&&\lambda_{1}>0
\label{eq:KKTlambda_1}\\
&&\eta_2>0
\label{eq:KKTeta_1}\\
&&\tr\left(\Wbf_2\Sbf_2^*\right)=0
\label{eq:KKTWS_1}
\\
&&\Wbf_2\succeq\0bf
\label{eq:KKTW0_1}
\esq
where $\Sbf^*$ is given in (\ref{eq:SmisoZss1}) which implies
\bqa
\phi^*=\frac{\pi}{2}.
\eqa
We note that since the constraint $R_1+R_2\leq g_{s2}$ in (\ref{eq:MIMOsuml}) is removed, the associated Lagrangian multiplier $\lambda_2$ in (\ref{eq:KKTrateI})-(\ref{eq:KKTW0}) is also removed (which is equivalent to setting $\lambda_2=0$).

Solving (\ref{eq:KKTrateSl_1})-(\ref{eq:KKTW0_1}), we have
\bqa
\Wbf_2=\left[\begin{array}{cc}
               0 &\quad 0 \\
               0 &\quad \dfrac{a}{2(1+P_1+aP_2)}
             \end{array}
\right].
\eqa
By Proposition \ref{prop:misoSC}, (\ref{eq:MISOZS1}) is the sum-rate capacity if
\bqa
&&\cos^2\theta < a\\
&&\Wbf_2\succeq\frac{1}{2}\cdot\frac{a\sin^2\theta}{a-\cos^2\theta}\left[\begin{array}{cc}
    0 &\quad 0 \\
    0 &\quad 1
                                                                        \end{array}
\right].
\eqa
The above two conditions reduce to (\ref{eq:MISOZScond1_2}). We note that $\cos^2\theta < a$ is redundant since (\ref{eq:MISOZScond1_2}) implies $a>1$.

Next, we prove the sum-rate capacity for conditions (\ref{eq:MISOZScond2_1}) and (\ref{eq:MISOZScond2_2}). By Lemma \ref{lemma:MISOZsl}, the maximal sum rate of (\ref{eq:MISOZhk}) is (\ref{eq:MISOZS2}) and is achieved by (\ref{eq:SmisoZss2}) which implies
\bqa
\phi^*=\phi_{ez}.
\eqa
There exist Lagrangian multipliers that satisfy
\bsq
&&\gamma+\lambda_{1}=1
\label{eq:KKTrateSl_2}\\
&&\Wbf_2=-\frac{\gamma\hp\hp^T}{2\left(1+P_2\sin^2\left(\theta+\phi_{ez}\right)\right)}
    -\frac{\lambda_{1}\fp\fp^T}{2\left(1+P_1+aP_2\sin^2\phi_{ez}\right)}+\eta_2\Ibf
\label{eq:KKTW2_2}\\
&&\gamma>0
\label{eq:gamma_2}\\
&&\lambda_{1}>0
\label{eq:KKTlambda_2}\\
&&\eta_2>0
\label{eq:KKTeta_2}\\
&&\tr\left(\Wbf_2\Sbf_2^*\right)=0
\label{eq:KKTWS_2}
\\
&&\Wbf_2\succeq\0bf.
\label{eq:KKTW0_2}
\esq
We note that we also removed the terms associate with $\lambda_2$ from (\ref{eq:KKTrateI})-(\ref{eq:KKTW0}) for the same reason. By solving (\ref{eq:KKTrateSl_2})-(\ref{eq:KKTW0_2}), we have
\bqa
\lambda_1&{}={}&\frac{\sin2\left(\theta+\phi_{ez}\right)}{\sin2\left(\theta+\phi_{ez}\right)-\dfrac{a\sin2\phi_{ez}}{1+P_1}}\\
\gamma&{}={}&1-\lambda_1\\
\eta_2&{}={}&k\sin^2\phi_{ez}+\frac{(1-\lambda_1)\sin^2\theta}{2\left(1+P_2\sin^2\left(\theta+\phi_{ez}\right)\right)}\\
\Wbf_2&{}={}&k\left[\begin{array}{cc}
                      \cos^2\phi_{ez} &\quad -\sin\phi_{ez}\cos\phi_{ez} \\
                       -\sin\phi_{ez}\cos\phi_{ez} & \quad \sin^2\phi_{ez}
                    \end{array}
\right]
\eqa
where
\bqa
k=\frac{1-\lambda_1}{2\left(1+P_2\sin^2\left(\theta+\phi_{ez}\right)\right)}\cdot\frac{\sin2\theta}{\sin2\phi_{ez}}.
\eqa
We note that under condition (\ref{eq:MISOZScond1_2}), we have $\frac{\pi}{2}<\phi_{ez}+\theta<\pi$.

By Proposition \ref{prop:misoSC}, (\ref{eq:MISOZS1}) is the sum-rate capacity if
\bqa
&&\sin^2\left(\theta+\phi_{ez}\right)< a\sin^2\phi_{ez}
\label{eq:misoZS2cond1}\\
&&k\geq\frac{\lambda_1}{2}\cdot\frac{a\sin^2\theta}{a\sin^2\phi_{ez}-\sin^2\left(\theta+\phi_{ez}\right)}.
\label{eq:misoZS2cond2}
\eqa
Condition (\ref{eq:misoZS2cond1}) is satisfied by (\ref{eq:phiStar}), and condition (\ref{eq:misoZS2cond2}) is satisfied by (\ref{eq:MISOZScond2_2}).
\epf

The MISO ZIC under conditions (\ref{eq:misozsCond0}), or (\ref{eq:MISOZScond1_2}), or (\ref{eq:MISOZScond2_1}) and (\ref{eq:MISOZScond2_2}) is said to have the generally strong interference sum-rate capacity of type I (very strong interference), type II and type III, respectively.

\begin{proposition}
For the MISO ZIC defined in (\ref{eq:eqZ}), if $a\cos^2\theta\geq 1+P_1$, then the capacity region is
\bqa
&&0\leq R_1\leq \frac{1}{2}\log(1+P_1)\\
&&0\leq R_2\leq \frac{1}{2}\log(1+P_2)
\eqa
and is achieved by (\ref{eq:misoZvsS}). If $a\cos^2\theta\leq 1+P_1$, then for any $\phi$ that satisfies
\bqa
Q(\phi)\geq 0,\quad \phi\in\Phibf
\label{eq:MISOZcbcond}
\eqa
where
\bqa
\Phibf&{}\triangleq{}&\left\{\begin{array}{ll}
    \left[\tau\left(\frac{\pi}{2}-\theta\right),\frac{\pi}{2}\right] &\qquad\textrm{if }\frac{a}{1+P_1}\leq\cos^2\theta\leq\frac{1+P_1}{a} \\
    \left[\tau\left(\frac{\pi}{2}-\theta\right),\phi_{ez}\right] & \qquad\textrm{if }\cos^2\theta\leq\min\left\{\frac{a}{1+P_1},\frac{1+P_1}{a}\right\}
                             \end{array}
\right.\\
%
%
Q(\phi)&{}\triangleq{}&a\sin^2\phi-\sin^2\left(\theta+\tau\phi\right)+\frac{\sin2\left(\theta+\tau\phi\right)\sin^2\theta}{\sin2\theta}
    \left(1+P_1+aP_2\sin^2\phi\right)
\eqa
where $\tau=\sign\left(\cos\theta\right)$, the following rate pair is on the boundary of the capacity region:
\bqa
R_1&{}={}&\frac{1}{2}\log\left(1+P_1+aP_2\sin^2\phi\right)-R_2
\label{eq:misoZCr1}\\
R_2&{}={}&\frac{1}{2}\log\left(1+P_2\sin^2\left(\theta+\phi\right)\right);
\label{eq:misoZCr2}
\eqa
and the rate pair is achieved by choosing $\Sbf$ as (\ref{eq:misoZsS}) and jointly decoding the signal and the interference at receiver $1$.
\label{prop:misozB}
\end{proposition}
\bpf
We prove only the case $\cos\theta\geq 0$. Since when $a\cos^2\theta\geq 1+P_1$, the MISO ZIC has very  strong interference and the capacity region is trivially proved, we need to consider only the case with $a\cos^2\theta\leq 1+P_1$. By Lemma \ref{lemma:MISOZIB}, the rate pair $(R_1,R_2)$ in (\ref{eq:misoZCr1}) and (\ref{eq:misoZCr2}) is on the boundary of the inner bound defined in (\ref{eq:MISOZiB1}). Therefore, this $R_1$ given in (\ref{eq:misoZCr1}) is the maximum in the optimization problem (\ref{eq:MIMOIB}) with $r=R_2$ given in (\ref{eq:misoZCr2}). By Lemma \ref{lemma:MIMOIB}, there exist Lagrangian multipliers that satisfy
\bsq
&&\alpha_1+\beta_1=1
\label{eq:KKTrateI_1}\\
&&\Kbf_2=-\frac{\alpha_2\hp\hp^T}{2\left(1+P_2\sin^2(\theta+\phi)\right)}-\frac{\beta_1\fp\fp^T}{2\left(1+P_1+aP_2\sin^2\phi\right)}+\nu_2\Ibf\\
&&\alpha_1=0\\
&&\alpha_2>0\\
&&\beta_1>0\\
&&\nu_2>0\\
&&\tr\left(\Kbf_2\Sbf_2^*\right)=0\\
&&\Kbf_2\succeq\0bf.
\label{eq:KKTK0_1}
\esq
We note that $\alpha_1=0$ simply because constraint $R_1\leq\frac{1}{2}\log(1+P_1)$ is always inactive by $\frac{\pi}{2}-\theta\leq\phi\leq\frac{\pi}{2}$ and $a\leq (1+P_1)\cos^2\theta$. Then we have
\bsq
\alpha_2&{}={}&-\frac{a\left(1+P_2\sin^2(\theta+\phi)\right)\sin2\phi}{\left(1+P_1+aP_2\sin^2\phi\right)\sin2(\theta+\phi)}\\
\beta_1&{}={}&1\\
\nu_2&{}={}&k\sin^2\phi+\frac{\alpha_2\sin^2\theta}{2\left(1+P_2\sin^2(\theta+\phi)\right)}\\
\Wbf_2&{}={}&k\left[\begin{array}{cc}
                      \sin^2\phi &\quad -\sin\phi\cos\phi \\
                      -\sin\phi\cos\phi &\quad \cos^2\phi
                    \end{array}
\right]
\esq
where
\bqa
k=\frac{\alpha_2\sin2\theta}{2\left(1+P_2\sin^2(\theta+\phi)\right)\sin2\phi}.
\eqa
We note that $\alpha_2>0$ since $\pi\leq\sin2(\theta+\phi)\leq\frac{3}{2}\pi$.

By Proposition \ref{prop:misoCB}, this $(R_1^*(r),r)$ is on the boundary of the capacity region if 
\bqa
&&{\sin^2(\theta+\phi)}<{a\sin^2\phi}
\label{eq:cond1}\\
&&k\geq\frac{1}{2}\frac{a\sin^2\theta}{a\sin^2\phi-\sin^2\left(\theta+\phi\right)}.
\label{eq:cond2}
\eqa
Condition (\ref{eq:cond2}) is equivalent to $Q(\phi)\geq 0$ in (\ref{eq:MISOZcbcond}). Condition (\ref{eq:cond1}) is satisfied by requiring $Q(\phi)\geq 0$, since the third term of $Q(\phi)$ is always non-positive.
\epf
\begin{remark}
Propositions \ref{prop:misozS} and \ref{prop:misozB} establish the full capacity region of a MISO ZIC with generally strong interference. When $\frac{a}{1+P_1}\leq\cos^2\theta\leq\frac{1+P_1}{a}$ and $Q(\phi)\geq 0$ for all $\phi\in\Phibf$, the capacity boundary points consist of (see Fig. \ref{fig:full1} as an example)
\bqa
\hspace{-.3in}\left\{\begin{array}{ll}
         \left\{\begin{array}{l}
                  R_1=\frac{1}{2}\log\left(1+P_1+aP_2\sin^2\phi\right)-R_2 \\
                  R_2=\frac{1}{2}\log\left(1+P_2\sin^2\left(\theta+\phi\right)\right)
                \end{array}
         \right\},  &\quad\phi\in\Phibf \\
         R_1+R_2=\frac{1}{2}\log(1+P_1+aP_2), &\quad  \frac{1}{2}\log\left(1+\frac{aP_2}{1+P_1}\right)  \leq R_2\leq\frac{1}{2}\log\left(1+P_2\cos^2\theta\right)\\
         R_1=\frac{1}{2}\log(1+P_1),&\quad 0\leq R_2\leq\frac{1}{2}\log\left(1+\frac{aP_2}{1+P_1}\right)\\
         R_2=\frac{1}{2}\log(1+P_2),&\quad 0\leq R_1\leq\frac{1}{2}\log\left(1+P_1+aP_2\cos^2\theta\right)-R_2.
       \end{array}
\right.
\eqa
When $\cos^2\theta\leq\min\left\{\frac{a}{1+P_1},\frac{1+P_1}{a}\right\}$ and $Q(\phi)\geq 0$ for all $\phi\in\Phibf$, the capacity boundary points consist of (see Fig. \ref{fig:full2} as an example)
\bqa
\left\{\begin{array}{ll}
         \left\{\begin{array}{l}
                  R_1=\frac{1}{2}\log\left(1+P_1+aP_2\sin^2\phi\right)-R_2 \\
                  R_2=\frac{1}{2}\log\left(1+P_2\sin^2\left(\theta+\phi\right)\right)
                \end{array}
         \right\},  &\quad\phi\in\Phibf\\
         R_1=\frac{1}{2}\log(1+P_1),&\quad 0\leq R_2\leq\frac{1}{2}\log\left(1+\frac{aP_2}{1+P_1}\right)\\
         R_2=\frac{1}{2}\log(1+P_2),&\quad 0\leq R_1\leq\frac{1}{2}\log\left(1+P_1+aP_2\cos^2\theta\right)-R_2.
       \end{array}
\right.
\eqa
\end{remark}

\subsection{Symmetric MISO IC}

A symmetric MISO IC has $\theta_1=\theta_2=\theta\in\left(0,\frac{\pi}{2}\right)$, $a_1=a_2=a>0$ and $P_1=P_2=P>0$. In this section, we derive sufficient conditions to determine the sum-rate capacity with generally strong interference. The derivation is similar to that of the MISO ZIC and is hence omitted. In the following, we only summarize the main result.

By symmetry, the maximal sum rate of region (\ref{eq:misoI}) is determined by
\bqa
\max&&\quad R_1+R_2\nn\\
\textrm{subject to}&&\quad R_1+R_2\leq q_u(\phi)\nn\\
&&\quad R_1+R_2\leq q_s(\phi)\nn\\
&&\quad 0\leq\phi\leq\frac{\pi}{2}
\label{eq:misoSlsymm}
\eqa
where
\bqa
q_u(\phi)&{}={}&\log\left(1+P\sin^2(\theta+\phi)\right)\\
q_s(\phi)&{}={}&\frac{1}{2}\log\left(1+P\sin^2(\theta+\phi)+aP\sin^2\phi\right).
\eqa
Obviously
\bqa
\max_{\phi\in\left[0,\frac{\pi}{2}\right]}q_u(\phi)=q_u\left(\phi=\phi_u\right)=\log(1+P)
\eqa
where
\bqa
\phi_u=\frac{\pi}{2}-\theta.
\eqa
It can be shown that
\bqa
\max_{\phi\in\left[0,\frac{\pi}{2}\right]}q_s(\phi)=q_s\left(\phi=\phi_s\right)
\eqa
where
\begin{subequations}
\begin{numcases}{\phi_s=}
\frac{\pi}{2}-\frac{1}{2}\textrm{atan}\left(\frac{\sin2\theta}{a+\cos2\theta}\right),\quad \textrm{if } a+\cos2\theta>0\\
\frac{\pi}{4}, \hspace{1.65in}\textrm{if } a+\cos2\theta=0\\
-\frac{1}{2}\textrm{atan}\left(\frac{\sin2\theta}{a+\cos2\theta}\right),\hspace{.35in} \textrm{if } a+\cos2\theta<0.
\end{numcases}
\end{subequations}
Define the set
\bqa
\Phi_e\triangleq\left\{\phi\left|q_u(\phi)=q_s(\phi),0\leq\phi\leq\frac{\pi}{2}\right.\right\}
\eqa
and denote
\bqa
\phi_e=\arg\max_{\phi\in\Phi_e} q_u(\phi).
\eqa
The maximum in problem (\ref{eq:misoSlsymm}) and the corresponding optimal $\phi^*$ are given in Tab. \ref{tab:misosymm}.
\begin{table}[htp]\caption{\label{tab:misosymm} The solution of problem (\ref{eq:misoSlsymm}) for a symmetric MISO IC.}
{\centerline{\begin{tabular}{|c|c|c|c|c|}
  \hline
    case & condition & maximum & $\phi^*$ & active constraints  \\
  \hline
    I & $q_s\left(\phi_u\right)\geq q_u\left(\phi_u\right)$ & $\log(1+P)$ & $\phi_u$ & $q_u$ \\
  \hline
    II & $q_s\left(\phi_s\right)\leq q_u\left(\phi_s\right)$ & $\frac{1}{2}\log(1+P\sin^2(\theta+\phi_s)+aP\sin^2\phi_s)$ & $\phi_s$ & $q_s$ \\
  \hline
    III & $\begin{array}{c}
             q_s\left(\phi_u\right)< q_u\left(\phi_u\right) \\
             q_s\left(\phi_s\right)> q_u\left(\phi_s\right)
           \end{array}
    $ & $\frac{1}{2}\log(1+P\sin^2(\theta+\phi_e)+aP\sin^2\phi_e)$ & $\phi_e$ & $q_u$, $q_s$ \\
  \hline
\end{tabular}}}
\end{table}

By symmetry, the Lagrangian multipliers in (\ref{eq:KKTrateSl})-(\ref{eq:KKTW0}) satisfy $\lambda_1=\lambda_2\triangleq\lambda$, $\eta_1=\eta_2\triangleq\eta$ and $\Wbf_1=\Wbf_2\triangleq\Wbf$. Using Tab. \ref{tab:misosymm} and Proposition \ref{prop:misoSC}, we obtain sufficient conditions for a symmetric MISO IC to have generally strong interference:

\be
\item Case I: the constraint $q_s$ is inactive and thus the Lagrangian multiplier associated with this constraint is $\lambda=0$. By Proposition \ref{prop:misoSC}, $\log(1+P)$ is the sum-rate capacity. In this case, the MISO IC has very strong interference.
\item Case II: the constraint $q_s$ is inactive and thus the Lagrangian multiplier associated with this constraint is $\gamma=0$. By solving (\ref{eq:KKTrateSl})-(\ref{eq:KKTW0}), we have
    \bqa
    \lambda&{}={}&\frac{1}{2}\\
    \Wbf&{}={}&k
    \left[\begin{array}{cc}
            \cos^2\phi_s &\quad -\cos\phi_s\sin\phi_s \\
            -\cos\phi_s\sin\phi_s &\quad \sin^2\phi_s
          \end{array}
    \right]
    \eqa
    where
    \bqa
    k=\frac{\sin\theta\cos\theta}{4\sin\phi\cos\phi\left(1+P\sin^2(\theta+\phi_s)+aP\sin^2\phi_s\right)}.
    \eqa
    By Proposition \ref{prop:misoSC}, if
    \bqa
    &&\sin^2\left(\theta+\phi_s\right)< a\sin^2\phi_s\label{eq:symmS2_1}\\
    &&k\geq\frac{1}{4}\cdot\frac{a\sin^2\theta}{a\sin^2\phi_s-\sin^2(\theta+\phi_s)}\label{eq:symmS2_2}
    \eqa
    then the sum rate capacity is $q_s(\phi_s)$.
\item Case III: constraints $q_u$ and $q_s$ are both active, therefore, $\gamma\neq 0$ and $\lambda\neq 0$. By solving (\ref{eq:KKTrateSl})-(\ref{eq:KKTW0}), we have
    \bqa
    \lambda&{}={}&\frac{1}{d+2}\\
    \Wbf&{}={}&k
        \left[\begin{array}{cc}
            \cos^2\phi_e &\quad -\cos\phi_e\sin\phi_e \\
            -\cos\phi_e\sin\phi_e &\quad \sin^2\phi_e
          \end{array}
    \right]\nn\\
    \eqa
    where
    \bqa
    k&{}={}&\frac{\left(d\left(1+P\sin^2\left(\theta+\phi_e\right)\right)+1\right)\sin\theta\cos\theta}
        {2(d+2)\left(1+P\sin^2\left(\theta+\phi_e\right)+aP\sin^2\phi_e\right)\sin\phi_e\cos\phi_e}\\
    d&{}={}&-\frac{\sin2\left(\theta+\phi_e\right)+a\sin2\phi_e}{\left(1+P\sin^2\left(\theta+\phi_e\right)\right)
        \sin2\left(\theta+\phi_e\right)}.
    \eqa
    By Proposition \ref{prop:misoSC}, if
    \bqa
    &&\sin^2\left(\theta+\phi_e\right)< a\sin^2\phi_e\label{eq:symmS3_1}\\
    &&k\geq\frac{\lambda}{2}\cdot\frac{a\sin^2\theta}{a\sin^2\phi_e-\sin^2(\theta+\phi_e)}\label{eq:symmS3_2}
    \eqa
    then $q_u(\phi_e)$ (or $q_s(\phi_e)$ is the sum-rate capacity.
\ee
\section{Numerical examples}
\label{sec:example}
\begin{example}
Consider a MIMO IC with
\bqn
&&\Hbf_1=\left[\begin{array}{cc}
    1.1388&\quad   -0.2236\\
    0.8445&\quad   -2.7614
             \end{array}
\right],\quad
\Fbf_1=\left[\begin{array}{cc}
    0.1489&\quad    5.0975\\
    1.3055&\quad    1.9099
             \end{array}
\right]\\
&&\Hbf_2=\left[\begin{array}{cc}
    1.1307&\quad    1.0983\\
    0.1415&\quad    0.2041
             \end{array}
\right],\quad
\Fbf_2=\left[\begin{array}{cc}
   -0.0970&\quad    0.7639\\
    1.9346&\quad    1.4774
             \end{array}
\right]\\
&&P_1=P_2=10.
\eqn
The maximal sum rate for the achievable region (\ref{eq:MIMOhk}) is
\bqn
R_1+R_2\leq\frac{1}{2}\log\left|\Ibf+\Hbf_1\Sbf_1^*\Hbf_1^T+\Fbf_2\Sbf_2^*\Fbf_2^T\right|= 3.2998
\eqn
and is achieved by
\bqn
\Sbf_1^*=\left[\begin{array}{cc}
    8.2319&\quad    0.3636\\
    0.3636&\quad    1.7681
             \end{array}
\right],\quad
\Sbf_2^*=\left[\begin{array}{cc}
    7.7370&\quad    4.1843\\
    4.1843&\quad    2.2630
             \end{array}
\right].
\eqn
The corresponding Lagrangian multipliers are
\bqn
&&\lambda_1=1,\quad \lambda_2=0,\quad \eta_1= 0.0545,\quad \eta_2=0.0394\\
&&\Wbf_1=\0bf,\quad \Wbf_2=\left[\begin{array}{cc}
    0.3794&\quad   -0.7015\\
   -0.7015&\quad    1.2972
             \end{array}
\right]\times 10^{-2}.
\eqn
Since only $\lambda_1>0$, the matrix $\Abf_2$ that satisfies (\ref{eq:Markov}) can be chosen as
\bqn
\Abf_2=\left[\begin{array}{cc}
    0.2802&\quad    0.5985\\
    0.1146&\quad    0.0789
             \end{array}
\right]\quad \textrm{and } \Abf_2\Abf_2^T\preceq\Ibf.
\eqn
From (\ref{eq:O}) we have $\Obf_2=\0bf$ and hence $\Wbf_2\succeq\lambda_1\Obf_2$.
By Theorem \ref{theorem:MIMOs}, this MIMO IC has generally strong interference. Therefore, the sum-rate capacity is $R_1+R_2=3.2998$.

We note that in this case since $\Sbf_1$ has full rank, the corresponding $\Abf_1$ that satisfies (\ref{eq:Markov}) has to be $\Abf_1=\Fbf_1^{-1}\Hbf_1$, and $\Abf_1\Abf_1^T\npreceq\Ibf$. Therefore, \cite[Proposition 3]{Shang-etal:10IT_mimo} does not apply, and this MIMO IC does not have strong interference in the sense of \cite{Costa&ElGamal:87IT}.
\end{example}
\begin{example}
Consider a symmetric MISO IC with $a=2$, $\theta=0.2\pi$ and $P=1$. By Tab. \ref{tab:misosymm}, this MISO IC satisfies the case III condition. The rate constraints $q_u(\phi)$ and $q_s(\phi)$ are shown in Fig. \ref{fig:symmS1}. The optimal input covariance matrix and the corresponding $\phi^*$ are
\bqn
\Sbf_1&{}={}&\Sbf_2=\left[\begin{array}{cc}
    0.8857 &\quad   0.3182\\
    0.3182 &\quad   0.1143
                    \end{array}
\right]\\
\phi^*&{}={}&\phi_e=0.3902\pi
\eqn
and the maximal sum rate is
\bqn
R_1+R_2=0.6532.
\eqn
The corresponding Lagrangian multipliers are
\bqn
\gamma=0.2627,\quad \lambda=0.3686,\quad \eta=0.1974, \quad \Wbf=0.1768\left[\begin{array}{cc}
    \cos^2\phi^* & \quad -\cos\phi^*\sin\phi^* \\
    -\cos\phi^*\sin\phi^*& \quad \sin^2\phi^*
                                                                             \end{array}
\right].
\eqn
The matrix $\lambda\Obf$ is
\bqn
\lambda\Obf=0.1499\left[\begin{array}{cc}
    \cos^2\phi^* & \quad -\cos\phi^*\sin\phi^* \\
    -\cos\phi^*\sin\phi^*& \quad \sin^2\phi^*
                                                                             \end{array}
\right].
\eqn
Therefore, $\Wbf\succeq\lambda\Obf$ and $R_1+R_2=0.6532$ is the sum-rate capacity.
\end{example}
\begin{example}
Consider a symmetric MISO IC with $a=2$, $\theta=0.1\pi$ and $P=4$. By Tab. \ref{tab:misosymm}, this MISO IC satisfies the case II condition. The rate constraints $q_u(\phi)$ and $q_s(\phi)$ are shown in Fig. \ref{fig:symmS2}. The optimal input covariance matrix and the corresponding $\phi^*$ are
\bqn
\Sbf_1&{}={}&\Sbf_2=\left[\begin{array}{cc}
    3.9576&\quad    0.4096\\
    0.4096&\quad    0.0424
                    \end{array}
\right]\\
\phi^*&{}={}&\phi_s=0.4672\pi
\eqn
and the maximal sum rate is
\bqn
R_1+R_2= 1.2724.
\eqn
The corresponding Lagrangian multipliers are
\bqn
\gamma=0,\quad \lambda=0.5000,\quad \eta=0.0576, \quad \Wbf= 0.0563\left[\begin{array}{cc}
    \cos^2\phi^* & \quad -\cos\phi^*\sin\phi^* \\
    -\cos\phi^*\sin\phi^*& \quad \sin^2\phi^*
                                                                             \end{array}
\right].
\eqn
The matrix $\lambda\Obf$ is
\bqn
\lambda\Obf=0.0467\left[\begin{array}{cc}
    \cos^2\phi^* & \quad -\cos\phi^*\sin\phi^* \\
    -\cos\phi^*\sin\phi^*& \quad \sin^2\phi^*
                                                                             \end{array}
\right].
\eqn
Therefore, $\Wbf\succeq\lambda\Obf$ and $R_1+R_2=1.2724$ is the sum-rate capacity.
\end{example}
\begin{example}
Consider a MISO ZIC with $a=6$, $\theta=0.2\pi$, $P_1=9$ and $P_2=3$. The function $Q(\phi)$ and the inner and outer bounds for the capacity region are shown in Figs. \ref{fig:partial_para} and \ref{fig:partial}, respectively. When $\phi\in\left[\frac{\pi}{2}-\theta,\phi_0\right]$ where $\phi_0=0.3748\pi$, we have $Q(\phi)\geq 0$. By Proposition \ref{prop:misozB}, the rate pairs given in (\ref{eq:misoZCr1}) and (\ref{eq:misoZCr2}) with $\phi\in\left[\frac{\pi}{2}-\theta,\phi_0\right]$ are on the boundary of the capacity region. Those points consist of the curve segment $\widehat{CB}$ of the Han and Kobayashi (HK) inner bound in Fig. \ref{fig:partial}, where point $C(0.8474,0.6931)$ is a corner point corresponding to $\phi=\frac{\pi}{2}-\theta$ and point $B(0.9442,0.6724)$ is corresponding to $\phi=\phi_0$. The HK inner bound is obtained by rate splitting and superposition coding. Another inner bound is obtained by jointly decoding the signal and the interference. The two outer bounds are obtained using Lemma \ref{lemma:MISOZo}. In (\ref{eq:MISOZo}), the outer bound tight at point $C$ has $A=0.5046$ and the outer bound tight at point $B$ has $A=0.4298$.
\label{example:partial}
\end{example}
\begin{example}
Consider a MISO ZIC with $a=1.2$, $\theta=0.1\pi$, $P_1=0.5$ and $P_2=0.5$. The function $Q(\phi)$ and the capacity region are shown in Figs. \ref{fig:full1_para} and \ref{fig:full1}, respectively. Since this channel satisfies $\frac{a}{1+P_1}\leq\cos^2\theta\leq\frac{1+P_1}{a}$ and for all $\phi\in\left[\frac{\pi}{2}-\theta,\frac{\pi}{2}\right]$ we have $Q(\phi)\geq 0$, by Proposition \ref{prop:misozB}, the rate pairs given in (\ref{eq:misoZCr1}) and (\ref{eq:misoZCr2}) with $\phi\in\left[\frac{\pi}{2}-\theta,\frac{\pi}{2}\right]$ are on the boundary of the capacity region. Those points consist of the curve segment $\widehat{C_1B}$ in Fig. \ref{fig:partial}, where point $C_1(0.1544,0.2027)$ is a corner point determined by $\phi=\frac{\pi}{2}-\theta$ and point $B(0.1844,0.1866)$ is determined by $\phi=\frac{\pi}{2}$. By Proposition \ref{prop:misozS}, the rate pair at point $B$ also achieves the sum-rate capacity. Therefore, the line segment $\overline{BC_2}$ satisfying $R_1+R_2=0.3710$ is also the boundary of the capacity region where $C_2(0.2027,0.1682)$ is another corner point. Therefore, the entire capacity region is determined and the MISO ZIC has generally strong interference for the entire capacity region.
\end{example}
\begin{example}
Consider a MISO ZIC with $a=2$, $\theta=0.2\pi$, $P_1=2$ and $P_2=0.4$. The function $Q(\phi)$ and the capacity region are shown in Figs. \ref{fig:full2_para} and \ref{fig:full2}, respectively. Since this channel satisfies $\cos^2\theta\leq\min\left\{\frac{a}{1+P_1},\frac{1+P_1}{a}\right\}$ and for all $\phi\in\left[\frac{\pi}{2}-\theta,\phi_{ez}\right]$ where $\phi_{ez}=0.4959\pi$, we have $Q(\phi)\geq 0$, and by Proposition \ref{prop:misozB}, the rate pairs given in (\ref{eq:misoZCr1}) and (\ref{eq:misoZCr2}) with $\phi\in\left[\frac{\pi}{2}-\theta,\phi_{ez}\right]$ are on the boundary of the capacity region. Those points consist of the curve segment $\widehat{C_1C_2}$ in Fig. \ref{fig:partial}, where point $C_1( 0.4615,0.1682)$ is a corner point determined by $\phi=\frac{\pi}{2}-\theta$ and point $C_2(0.5493,0.1182)$ is another corner point determined by $\phi=\phi_{ez}$. By Proposition \ref{prop:misozS}, the rate pair at point $C_2$ also achieves the sum-rate capacity $R_1+R_2= 0.6675$. Therefore, the entire capacity region is determined and the MISO ZIC has generally strong interference for the entire capacity region.
\end{example}
\begin{example}
Fig. \ref{fig:AZIC} shows the maximal value of $a$ for a MISO ZIC to have generally strong interference sum-rate capacity, and the minimal value of $a$ for a MISO ZIC to have noisy interference sum-rate capacity \cite{Shang&Poor:11IT_submission_MIMO}. For all the MISO ZICs with $a$ and $\theta$ above the `Minimum of $a$ for GS IF' curve,  the sum-rate capacity is achieved by jointly decoding the signal and the interference. For all the MISO ZICs with $a$ and $\theta$ below the `Maximum of $a$ for NIF' curve,  the sum-rate capacity is achieved by treating interference as noise. We also show the region for the MISO ZIC to have the generally strong interference sum-rate capacity of case II (see eq. (\ref{eq:MISOZScond1_2})), and case III (see eqs. (\ref{eq:MISOZScond2_1}) and (\ref{eq:MISOZScond2_2})).
\end{example}
\begin{example}
Fig. \ref{fig:AICsymm} shows the maximal value of $a$ for a symmetric MISO IC to have generally strong interference sum-rate capacity, and the minimal value of $a$ for a MISO ZIC to have noisy interference sum-rate capacity \cite{Shang&Poor:11IT_submission_MIMO}. For all the symmetric MISO ICs with $a$ and $\theta$ above the `Minimum of $a$ for GS IF' curve,  the sum-rate capacity is achieved by jointly decoding the signal and the interference. For all the MISO ICs with $a$ and $\theta$ below the `Maximum of $a$ for NIF' curve,  the sum-rate capacity is achieved by treating interference as noise. We also show the region for the symmetric MISO IC to have the generally strong interference sum-rate capacity of case II (see eqs. (\ref{eq:symmS2_1}) and (\ref{eq:symmS2_2})), and case III (see eqs. (\ref{eq:symmS3_1}) and (\ref{eq:symmS3_2})).
\end{example}

\section{Conclusion}
In this paper, we have extended the capacity result for MIMO ICs with strong interference to those with generally strong interference. Although in both cases the capacity region is achieved by jointly decoding the signal and the interference, the strong interference conditions require the receivers to be able to decode the signal and interference for any input distribution, while for generally strong interference, the receivers are required to do so only for the capacity achieving input distributions. The generally strong interference conditions for a MIMO IC have been obtained and the application to SIMO and MISO ICs has also been discussed. The obtained conditions include existing capacity results for strong and very strong interference as special cases.

\appendix
\subsection{Proof of Lemma \ref{lemma:inner}}
\label{appendix:inner}
Generate a length-$n$ random vector $q^n$ with independent and identically distributed (i.i.d.) elements according to
\bqn
p\left(q^n\right)=\prod_{m=1}^np(q_m).
\eqn
Let $i$ denote the index of the messages transmitted by user $1$ and $i\in\left\{1,2,\cdots,e^{nR_1}\right\}$. For each $i$, generate a length-$n$ random vector $x_1^n$ with i.i.d. elements according to
\bqn
p\left(x_1^n\left|q^n\right.\right)=\prod_{m=1}^np\left(x_{1m}|q_m\right).
\eqn
We label this sequence as $x_1^n(i)$.

Let $j$ denote the index of the message transmitted by user $2$ and $j\in\left\{1,2,\cdots,e^{nR_2}\right\}$. For each $j$, generate a length-$n$ random vector $x_j^n$ with i.i.d. elements according to
\bqn
p\left(x_2^n\left|q^n\right.\right)=\prod_{m=1}^np\left(x_{2m}|q_m\right).
\eqn
We label this sequence as $x_2^n(j)$.

To send message indices $i$ to receivers $1$, transmitter $1$ sends the codeword $x_1^n(i)$. To send message index $j$ to receiver $2$, transmitter $2$ sends the codeword $x_2^n(j)$.

Receiver $1$ looks for unique indices $(\hat i,\hat j)$ such that
\bqa
\left(q^n,x_1^n\left(\hat i\right),x_2^n\left(\hat j\right),y_1^n\right)\in A_\epsilon^{(n)}\left(Q,X_1,X_2,Y_1\right)
\eqa
where $A_\epsilon^{(n)}$ denotes the set of jointly typical sequences.

Receiver $2$ looks for unique indices $\left(\hat i,\hat j\right)$ such that
\bqa
\left(q^n,x_1^n\left(\hat i\right),x_2^n\left(\hat j\right),y_2^n\right)\in A_\epsilon^{(n)}\left(Q,X_1,X_2,Y_2\right).
\eqa
An error occurs if there are no such indices or the indices are not unique.

By symmetry, we assume that the transmitted indices are $i=j=1$. For user $1$, we define the following event:
\bqa
E_{ij}^1=\left\{\left(q^n,x_1^n\left( i\right),x_2^n\left( j\right),y_1^n\right)\in A_\epsilon^{(n)}\left(Q,X_1,X_2,Y_1\right)\right\}.
\eqa
The error probability at receiver $1$ is
\bqa
P_{e1}&{}={}&\textrm{Pr}\left\{{E_{11}^1}^c\bigcup\cup_{(i\neq 1,\textrm{any }j)}E_{ij}^1\right\}\nn\\
&&\leq \textrm{Pr}\left\{{E_{11}^1}^c\right\}+\sum_{i\neq 1,j=1}\textrm{Pr}\left(E_{i1}^1\right)+\sum_{i\neq 1,j\neq 1}\textrm{Pr}\left(E_{ij}^1\right)\nn\\
&&\leq \epsilon+e^{n\left(R_1-I\left(X_1;Y_1|X_2Q\right)\right)}+e^{n\left(R_1+R_2-I\left(X_1X_2;Y_1|Q\right)\right)}.
\label{eq:pe1}
\eqa
Similarly, the error probability of receiver $2$ is
\bqa
P_{e2}\leq\epsilon+e^{n\left(R_2-I\left(X_2;Y_2|X_1Q\right)\right)}+e^{n\left(R_1+R_2-I\left(X_1X_2;Y_2|Q\right)\right)}.
\eqa
Therefore, the rate region in Lemma \ref{lemma:inner} is achievable.

\subsection{Proof of Lemma \ref{lemma:subset}}
\label{appendix:subset}
For completeness, we rewrite the simplified Han and Kobayashi region \cite{Chong-etal:08IT} in the following and denote it as $HK\left(W_1,W_2\right)$:
\bqn
0\leq R_1&{}\leq{}&I\left(X_1;Y_1|W_2Q\right)\\
0\leq R_2&{}\leq{}&I\left(X_2;Y_2|W_1Q\right)\\
R_1+R_2&{}\leq{}&I\left(X_1W_2; Y_1|Q\right) + I\left(X_2; Y_2|W_1W_2Q\right)\\
R_1+R_2&{}\leq{}&I\left(X_1;Y_1|W_1W_2Q\right) + I\left(X_2W_1;Y_2|Q\right)\\
R_1+R_2&{}\leq{}&I\left(X_1W_2;Y_1|W_1Q\right) + I\left(X_2W_1;Y_2|W_2Q\right)\\
2R_1+R_2&{}\leq{}&I\left(X_1W_2;Y_1|Q\right) + I\left(X_1;Y_1|W_1W_2Q\right)+I\left(X_2W_1;Y_2|W_2Q\right)\\
R_1+2R_2&{}\leq{}&I\left(X_2;Y_2|W_1W_2Q\right) + I\left(X_2W_1;Y_2|Q\right)+I\left(X_1W_2;Y_1|W_1Q\right).
\eqn
We denote the region defined in (\ref{eq:innerR1})-(\ref{eq:innerRs2}) as $\Rmat$. To show that $\Rmat$ is a subset of the Han and Kobayashi region, it suffices to show
\bqa
\Rmat\subseteq HK(X_1,X_2)\bigcup HK(\textrm{empty},X_2)\bigcup HK(X_1,\textrm{empty}).
\eqa
Let $\{R_1,R_2\}\in\Rmat$. Then $R_1$ and $R_2$ satisfy (\ref{eq:innerR1})-(\ref{eq:innerRs2}). If $R_1$ and $R_2$ also satisfy the extra constraint (\ref{eq:extra}), then $\{R_1,R_2\}\in HK(X_1,X_2)$. Otherwise, we have
\bqa
R_1+R_2>I\left(X_1;Y_2|X_2Q\right)+I\left(X_2;Y_1|X_1Q\right).
\eqa
We have only three possible scenarios:
\bqn
&&R_1\geq I\left(X_1;Y_2|X_2Q\right),\quad R_2\leq I\left(X_2;Y_1|X_1Q\right)\quad\textrm{or}\\
&&R_1\leq I\left(X_1;Y_2|X_2Q\right),\quad R_2\geq I\left(X_2;Y_1|X_1Q\right)\quad\textrm{or}\\
&&R_1\geq I\left(X_1;Y_2|X_2Q\right),\quad R_2\geq I\left(X_2;Y_1|X_1Q\right).
\eqn
Suppose $R_1\geq I\left(X_1;Y_2|X_2Q\right)$. By (\ref{eq:innerRs2}) we have
\bqa
R_2\leq I\left(X_2;Y_2|Q\right).
\label{eq:innerR2_2}
\eqa
On the other hand, $HK(\textrm{empty},X_2)$ is given by
\bqa
0\leq R_1&{}\leq{}& I\left(X_1;Y_1|X_2Q\right)\\
0\leq R_2&{}\leq{}& I\left(X_2;Y_2|Q\right)\\
R_1+R_2&{}\leq{}& I\left(X_1X_2;Y_1|Q\right)
\eqa
Since $\{R_1,R_2\}$ satisfies (\ref{eq:innerR1}), (\ref{eq:innerRs1}) and (\ref{eq:innerR2_2}), we have $\{R_1,R_2\}\in HK(\textrm{empty},X_2)$.
Similarly, if $R_2\geq I\left(X_2;Y_1|X_1Q\right)$, then $\{R_1,R_2\}\in HK(X_1,\textrm{empty})$.

\begin{figure*}[htp]
\centerline{\leavevmode \epsfxsize=5.5in \epsfysize=3.9in
\epsfbox{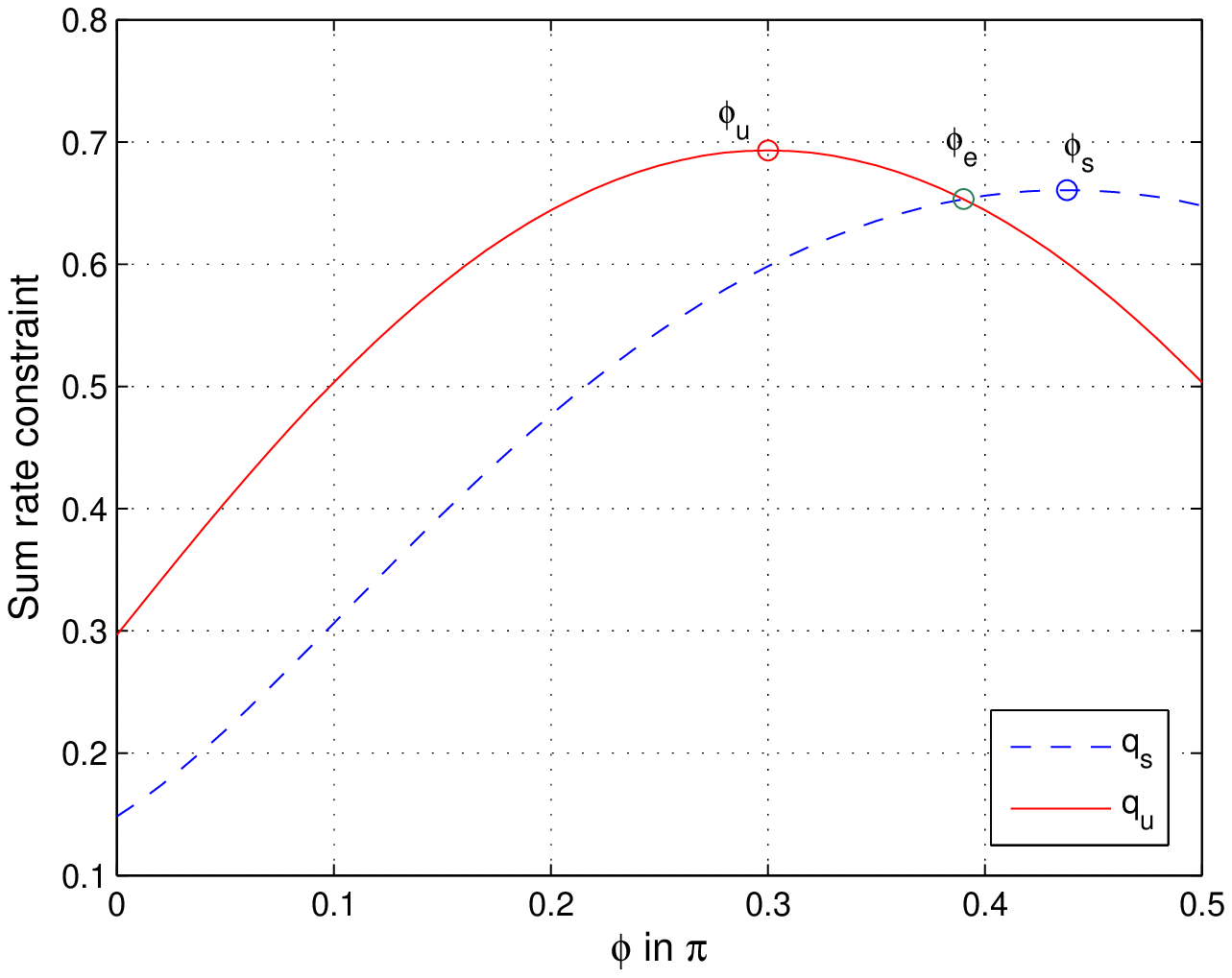}}\caption{$q_u(\phi)$ and $q_s(\phi)$ for a symmetric MISO IC with $a=2, \theta=0.2\pi, P=1$.}
\label{fig:symmS1}
\end{figure*}

\begin{figure*}[hbp]
\centerline{\leavevmode \epsfxsize=5.5in \epsfysize=3.78in
\epsfbox{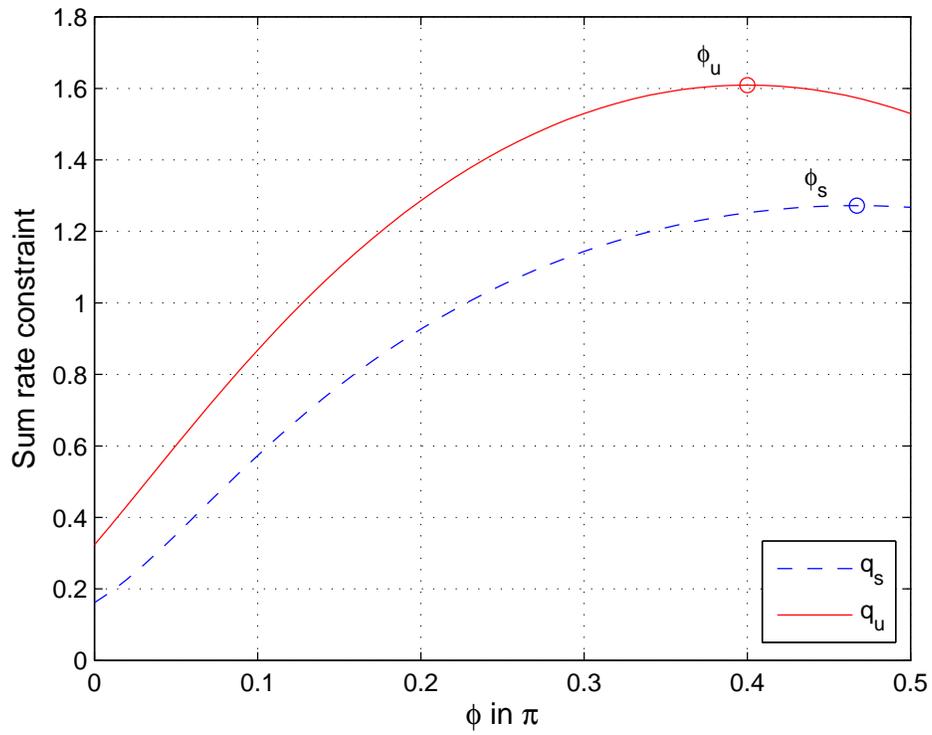}}\caption{$q_u(\phi)$ and $q_s(\phi)$ for a symmetric MISO IC with $a=2, \theta=0.1\pi, P=4$.}
\label{fig:symmS2}
\end{figure*}

\begin{figure*}[htp]
\centerline{\leavevmode \epsfxsize=5.5in \epsfysize=3.8in
\epsfbox{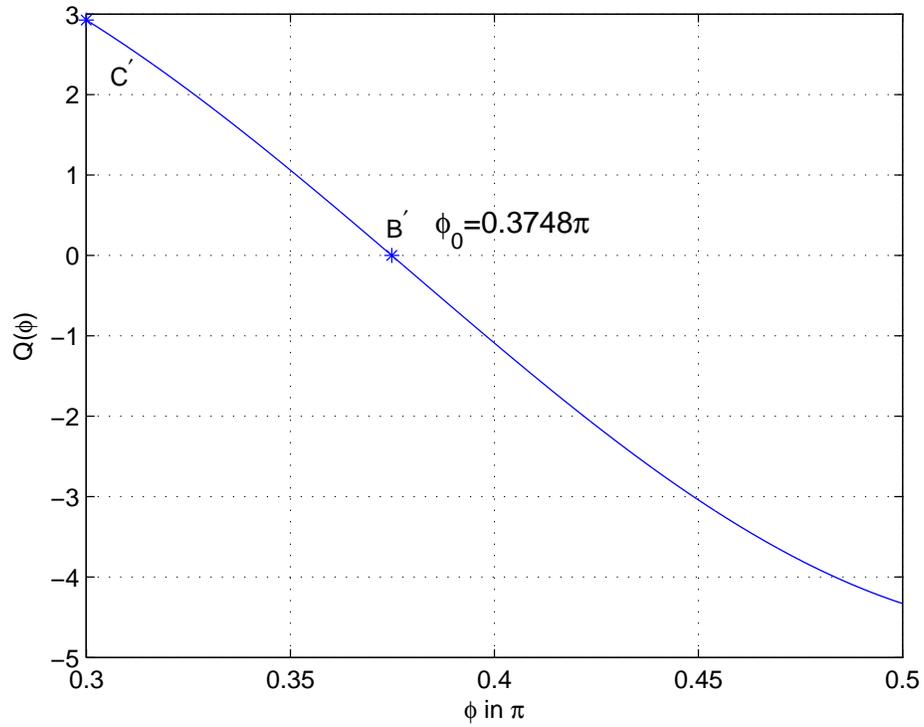}}\caption{$Q(\phi)$ with $\phi\in\left[\frac{\pi}{2}-\theta,\frac{\pi}{2}\right]$ for a MISO ZIC with $a=6, \theta=0.2\pi, P_1=9$ and $P_2=3$.}
\label{fig:partial_para}
\end{figure*}

\begin{figure*}[htp]
\centerline{\leavevmode \epsfxsize=5.5in \epsfysize=3.8in
\epsfbox{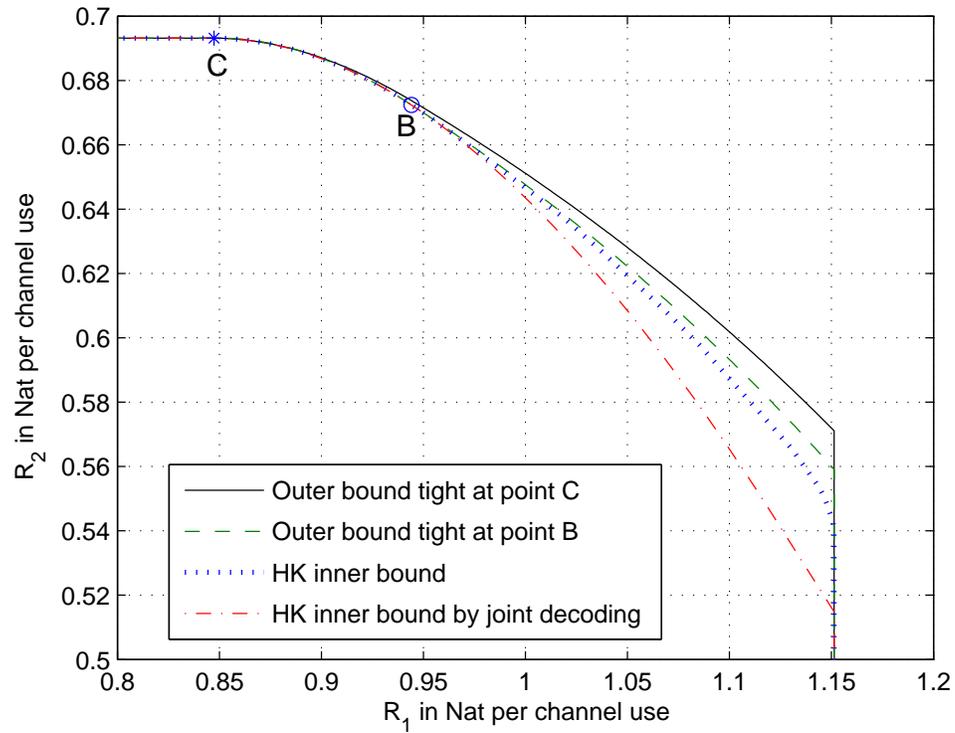}}\caption{Inner and outer bounds for the capacity region of a MISO ZIC with $a=6, \theta=0.2\pi, P_1=9$ and $P_2=3$.}
\label{fig:partial}
\end{figure*}

\begin{figure*}[htp]
\centerline{\leavevmode \epsfxsize=5.5in \epsfysize=3.8in
\epsfbox{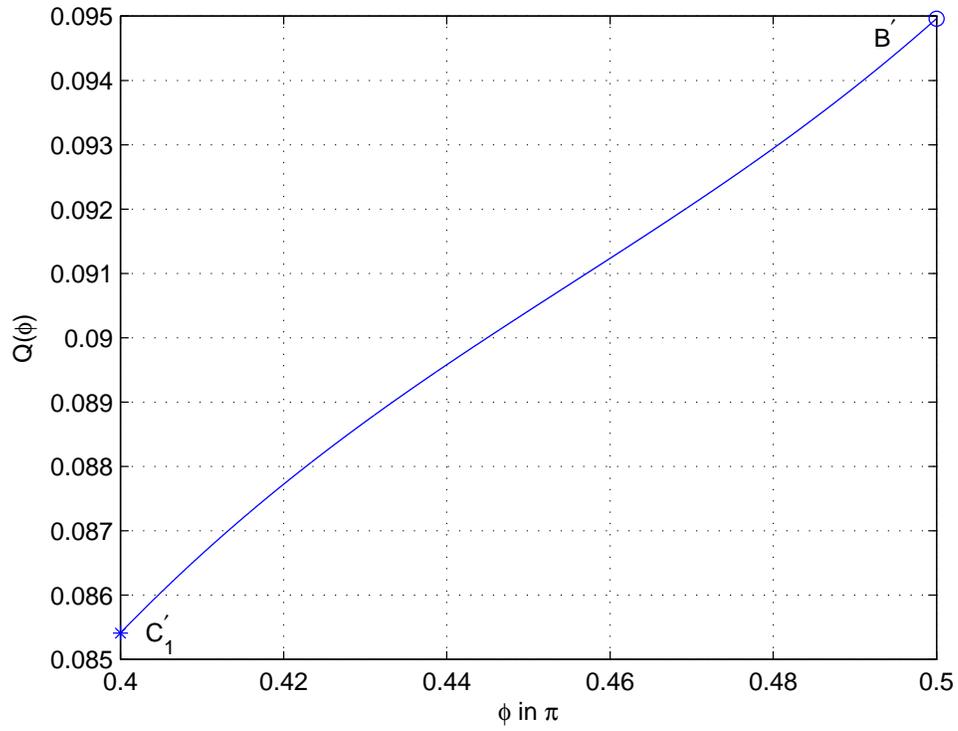}}\caption{$Q(\phi)$ with $\phi\in\left[\frac{\pi}{2}-\theta,\frac{\pi}{2}\right]$ for a MISO ZIC with $a=1.2, \theta=0.1\pi, P_1=0.5$ and $P_2=0.5$.}
\label{fig:full1_para}
\end{figure*}

\begin{figure*}[htp]
\centerline{\leavevmode \epsfxsize=5.5in \epsfysize=3.8in
\epsfbox{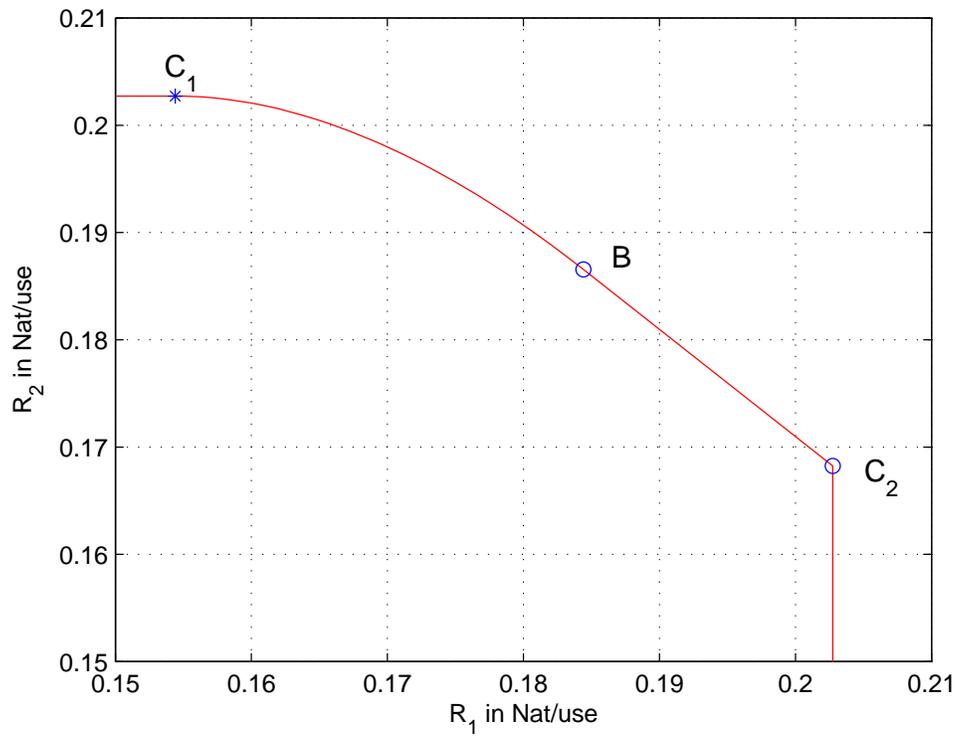}}\caption{The capacity region of a MISO ZIC with $a=1.2, \theta=0.1\pi, P_1=0.5$ and $P_2=0.5$.}
\label{fig:full1}
\end{figure*}

\begin{figure*}[htp]
\centerline{\leavevmode \epsfxsize=5.5in \epsfysize=3.8in 
\epsfbox{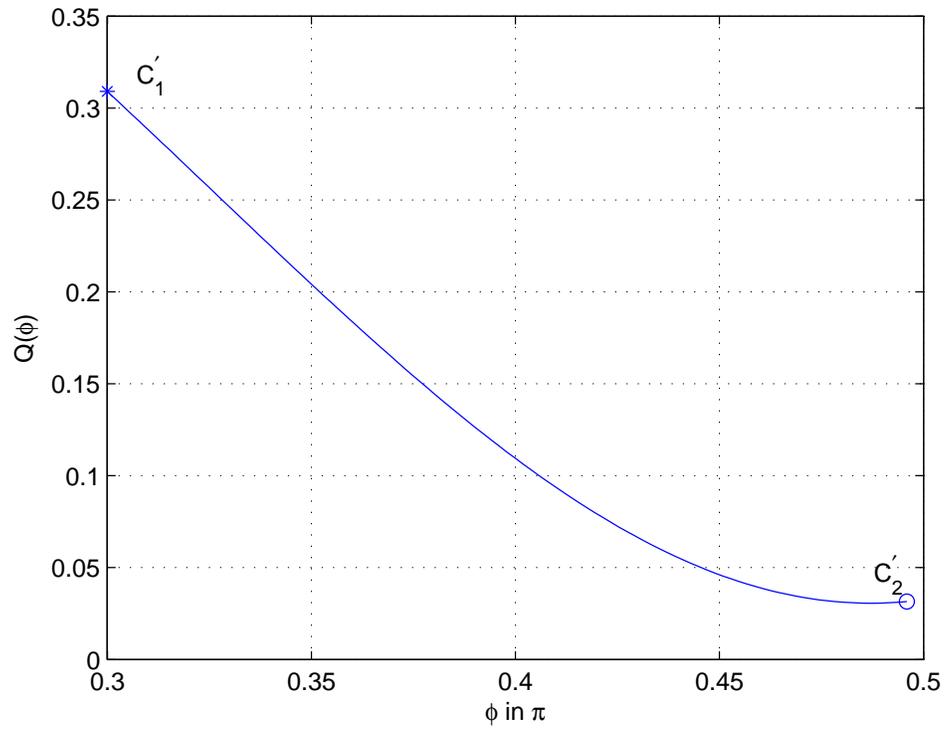}}\caption{$Q(\phi)$ with $\phi\in\left[\frac{\pi}{2}-\theta,\frac{\pi}{2}\right]$ for a MISO ZIC with $a=2, \theta=0.2\pi, P_1=2$ and $P_2=0.4$.}
\label{fig:full2_para}
\end{figure*}

\begin{figure*}[htp]
\centerline{\leavevmode \epsfxsize=5.5in \epsfysize=3.8in 
\epsfbox{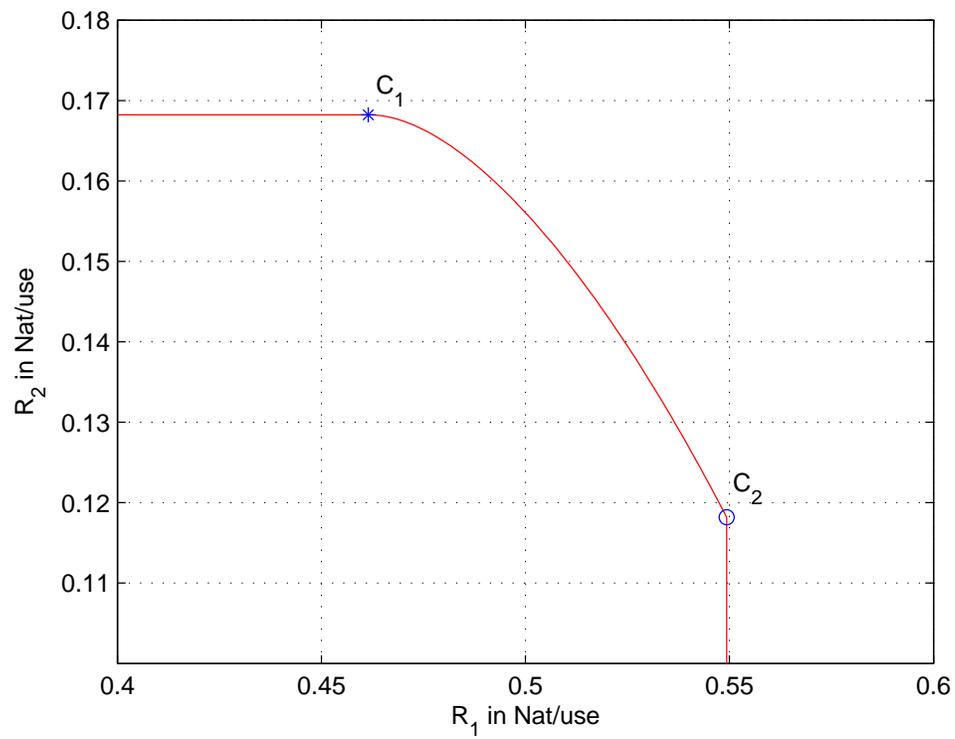}}\caption{The capacity region of a MISO ZIC with $a=2, \theta=0.2\pi, P_1=2$ and $P_2=0.4$.}
\label{fig:full2}
\end{figure*}

\begin{figure*}[htp]
\centerline{\leavevmode \epsfxsize=5.5in \epsfysize=3.8in 
\epsfbox{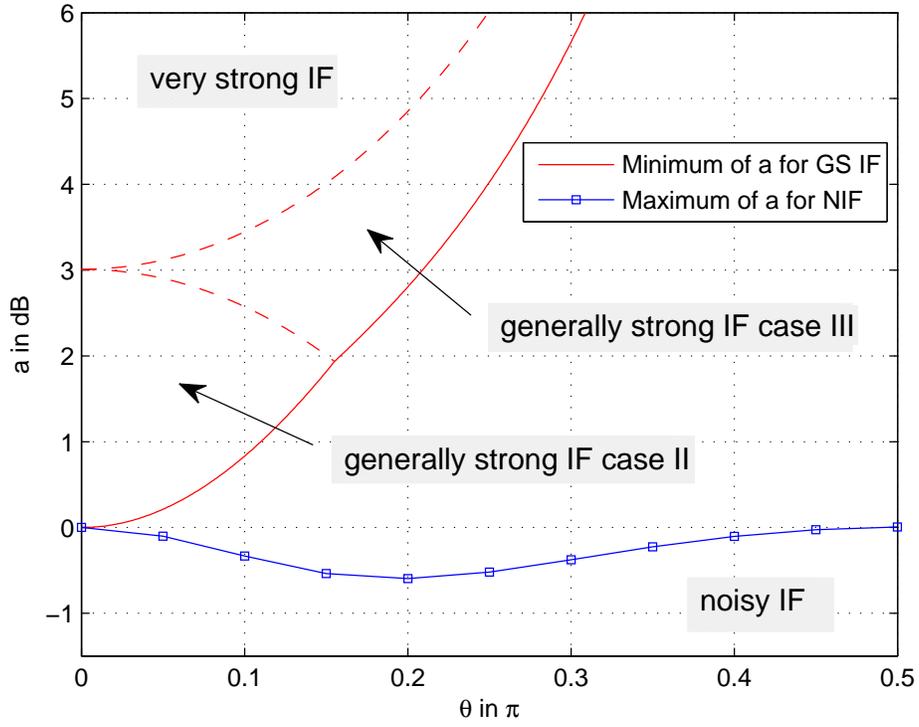}}\caption{The value of $a$ and the interference type for a MISO ZIC with $P_1=P_2=1$.}
\label{fig:AZIC}
\end{figure*}

\begin{figure*}[htp]
\centerline{\leavevmode \epsfxsize=5.5in \epsfysize=3.8in 
\epsfbox{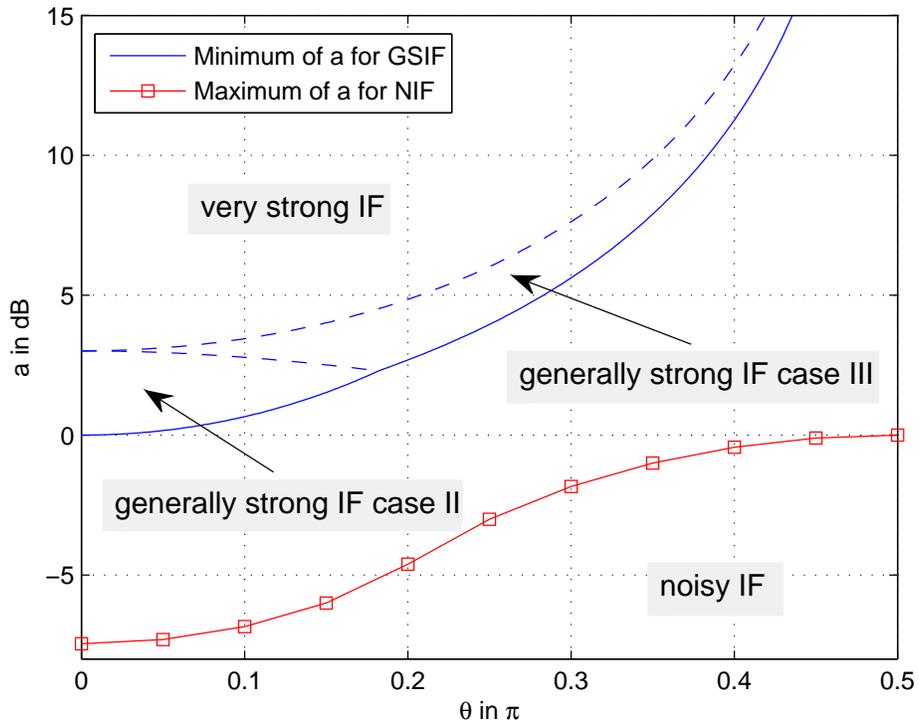}}\caption{The value of $a$ and the interference type for a symmetric MISO IC with $P=1$.}
\label{fig:AICsymm}
\end{figure*}

\bibliography{Journal,Conf,Misc,Book}
\bibliographystyle{IEEEbib}
\end{document}